\RequirePackage{fix-cm}
\RequirePackage{amsmath}
\documentclass[twocolumn]{svjour3}          
\smartqed  
\usepackage{graphicx}
\usepackage[acronym]{glossaries}
\usepackage{siunitx}
\usepackage{pgf,pgffor}
\usepackage[colorlinks=false]{hyperref}
\usepackage[misc]{ifsym}
\usepackage{amssymb}
\usepackage{bm}
\usepackage{mathtools}
\usepackage{tikz}
\usetikzlibrary{positioning}

\newacronym{NN}{NN}{neural network}
\newacronym{LHC}{LHC}{Large Hadron Collider}

\begin{document}

\title{Optimal statistical inference in the presence of systematic uncertainties using neural network optimization based on binned Poisson likelihoods with nuisance parameters}


\author{Stefan Wunsch \and Simon J\"orger \and Roger Wolf \and G\"unter Quast}


\institute{
Stefan Wunsch\textsuperscript{1,2} (corresponding author) \at
stefan.wunsch@cern.ch
\and
Simon J\"orger\textsuperscript{1} \at
simon.joerger@cern.ch
\and
Roger Wolf\textsuperscript{1} \at
roger.wolf@cern.ch
\and
G\"unter Quast\textsuperscript{1} \at
guenter.quast@kit.edu
\and
\textsuperscript{1} Karlsruhe Institute of Technology, Institute of Experimental Particle Physics, Karlsruhe, Germany\\
\textsuperscript{2} CERN, Geneva, Switzerland
}


\maketitle

\begin{abstract}
Data analysis in science, e.g., high-energy particle physics, is often subject to an intractable likelihood if the observables and observations span a high-dimensional input space. Typically the problem is solved by reducing the dimensionality using feature engineering and histograms, whereby the latter allows to build the likelihood using Poisson statistics. However, in the presence of systematic uncertainties represented by nuisance parameters in the likelihood, an optimal dimensionality reduction with a minimal loss of information about the parameters of interest is not known. This work presents a novel strategy to construct the dimensionality reduction with neural networks for feature engineering and a differential formulation of histograms so that the full workflow can be optimized with the result of the statistical inference, e.g., the variance of a parameter of interest, as objective. We discuss how this approach results in an estimate of the parameters of interest that is close to optimal and the applicability of the technique is demonstrated with a simple example based on pseudo-experiments and a more complex example from high-energy particle physics.
\end{abstract}

\glsresetall

\section{Introduction}
\label{sec:Introduction}

Measurements in many areas of research like, e.g., high-energy particle physics, are typically based on the statistical inference of one or more parameters of interest defined by the likelihood $\mathcal{L}(D,\bm{\theta})$ with the observables $\bm{x}\in X \subseteq \mathbb{R}^d$ building the dataset $D=\{\bm{x}_1, ..., \bm{x}_n\}\subseteq\mathbb{R}^{n\times d}$ and the parameters $\bm{\theta}$ of the statistical model. The likelihood would have to be evaluated for the dataset $D$ spanning a high-dimensional input space, which is computationally expensive and typically unfeasible. The dimension of $D$ can be reduced by the engineering of high-level observables and the usage of summary statistics. Analysts create high-level observables to reduce the dimension $d$ of a single observation to $k$, ideally without losing information about the parameters $\bm{\theta}$. An example from high-energy particle physics is the usage of the invariant mass of a decay system instead of the kinematic properties of all its constituents. The dimension $n$ of $D$ can be reduced with the computation of a summary statistic, for which histograms are frequently used so that the statistical model can be expressed in form of a likelihood, based on Poisson statistics. The dimension is thus reduced from the number of observations $n$ to the number of bins $h$ in the histogram, whereby the analyst tries to optimize the trade-off between a feasible number of bins and the loss of information about the parameters of interest. Applying both methods, the initial dimension of $D\subseteq\mathbb{R}^{n\times d}$ is reduced to $\mathbb{R}^{h\times k}$.

This paper discusses an analysis strategy using machine learning techniques, by which the suboptimal performance introduced by the reduction of dimensionality can be avoided resulting in estimates of the parameters of interest $\bm{\mu}\in\bm{\theta}$ close to optimal, e.g., with a minimal variance. We put emphasis on the applicability of this approach to analysis strategies as used for Higgs boson analyses at the \gls{LHC}~\cite{cowan2011asymptotic,atlas2011procedure,chatrchyan2012observation,aad2012observation}.

Section~\ref{sec:Method} presents the method in detail and section~\ref{sec:Related_work} puts the proposed technique in context of related work. Section~\ref{sec:Application_to_a_simple_example} shows the performance of the method with a simple example using pseudo-experiments of a two-component mixture model with signal and background and section~\ref{sec:Application_to_a_more_complex} applies the same approach to a more complex example from high-energy particle physics.

\section{Method}
\label{sec:Method}

The method requires as input an initial dataset $D \subseteq \mathbb{R}^{n\times d}$ used for the statistical inference of the parameters of interest with $n$ being the number of observations and $d$ the number of observables. To simplify the statistical evaluation, we typically reduce the number of observables by the engineering of high-level observables. Besides manual crafting of such features, a suited approach taken from machine learning is using a \gls{NN} function $\bm{f}(\bm{x}, \bm{\omega}) : \bm{x}\in X \subseteq \mathbb{R}^d \rightarrow \bm{f} \in F \subseteq \mathbb{R}^k$ with $\bm{\omega}$ being the free \gls{NN} parameters. After application of the \gls{NN}, we get a transformed dataset $D_\mathrm{NN} \subseteq \mathbb{R}^{n\times k}$ with $k$ the number of output nodes of the \gls{NN} architecture.

\begin{figure}
\centering
\includegraphics[width=0.8\linewidth]{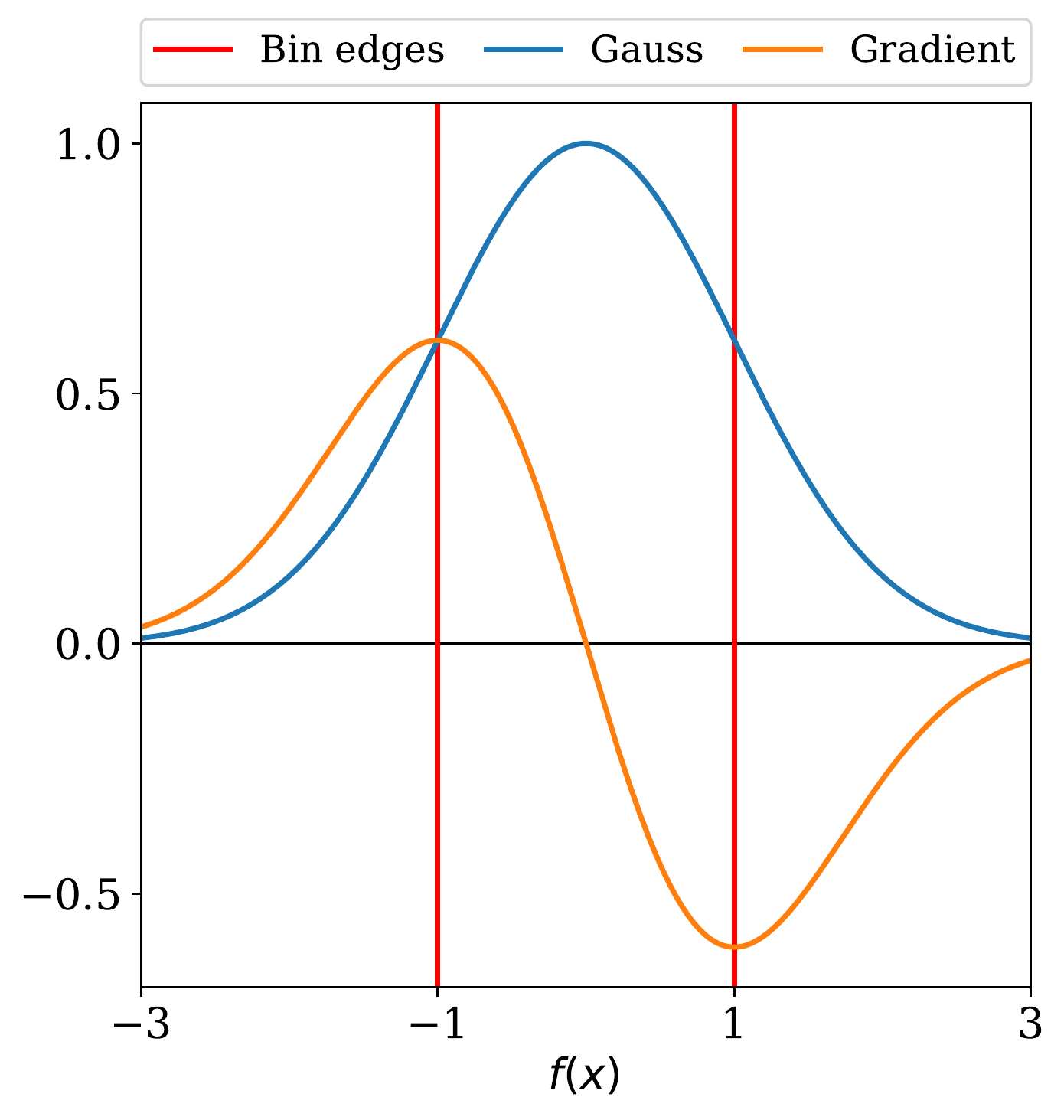}
\caption{The figure shows the approximation of the gradient of a single bin in a histogram with the gradient of a Gaussian $G$ normalized to $\max(G)=1$ with the standard deviation equal to the half-width of the bin.}
\label{fig:grad_bin}
\end{figure}

To reduce the dataset $D_\mathrm{NN}$ further, the number of observations $n$ is compressed using a histogram. Histograms are widely used as a summary statistic since counts follow the Poisson distribution and therefore are well suited to build the statistical model of the analysis. For example in high-energy particle physics, many statistical models and well established methods for describing the statistical model and systematic uncertainties are based on counts and bin\-ned Poisson likelihoods.
The resulting dataset is $D_\mathrm{H} \subseteq \mathbb{R}^{h\times k}$ using $h$ bins for the $k$-dimensional histogram. The count operation for a single bin in the histogram can be written as $C = \sum_{i=1}^n S\left(\bm{f}\left(\bm{x}_i,\bm{\omega}\right)\right)$ with
\begin{equation}
    S\left(\bm{f}\left(\bm{x}_i,\bm{\omega}\right)\right) = \begin{cases}
      1, &\text{if}\ \bm{f}\ \text{in the bin boundaries} \\
      0, &\text{otherwise}.
    \end{cases}
\end{equation}
In order to propagate the gradient from the result of the statistical inference to the free parameters $\bm{\omega}$ of the \gls{NN}, the histogram has to be differentiable. Since the derivative of $S$ is ill-defined on the edges of the bin and otherwise zero, the gradient is not suitable for optimization. Therefore, we use a smoothed approximation of the gradient~\cite{wunsch2019reducing} shown in figure~\ref{fig:grad_bin} for a one-dimensional bin. The approximation uses the similarity of $S$ to a Gaussian function $\mathcal{G}$ normalized to $\max(\mathcal{G}) = 1$ with the standard deviation being the half-width of the bin. We replace only the gradient of the operation $S$ and not the calculation of the count itself to keep the statistical model of the final analysis unchanged.

On top of the reduced dataset $D_\mathrm{H}$, we build the statistical model using a binned likelihood $\mathcal{L}(D_\mathrm{H},\bm{\theta})$ with $\bm{\theta}$ being the parameters of the statistical model. For a mixture model with the two processes signal $\bm{s}$ and background $\bm{b}$, the binned likelihood describing the statistical component is given by
\begin{equation}\label{eq:lh_stats}
    \mathcal{L}(D_\mathrm{H},\bm{\theta}) = \prod_{i=1}^h\mathcal{P}\left(d_i | \mu s_i + b_i\right)
\end{equation}
with $\mathcal{P}$ being the Poisson distribution, $\bm{d}$ the observation and $\mu \in \bm{\theta}$ the parameter of interest scaling the expectation of the signal process $\bm{s}$.

Moreover, the formulation of the statistical model allows to implement systematic uncertainties by adding nuisance parameters to the set of parameters $\bm{\theta}$. For the model in equation~\ref{eq:lh_stats}, a single nuisance parameter $\eta$ controlling a systematic variation $\bm{\Delta}$ of the expected bin contents results in
\begin{equation}\label{eq:lh_full}
    \mathcal{L}(D_\mathrm{H},\bm{\theta}) = \prod_{i=1}^h\mathcal{P}\left(d_i | \mu s_i + b_i + \eta\Delta_i\right) \,\mathcal{N}\left(\eta\right)
\end{equation}
with $\mathcal{N}$ being a standard normal distribution constraining the nuisance $\eta$. If the systematic variation is asymmetric, the additional nuisance term can be written as
\begin{equation}\label{eq:nuisance_asymmetric}
\max{(\eta, 0)}\bm{\Delta}_\mathrm{up} + \min{(\eta, 0)}\bm{\Delta}_\mathrm{down}
\end{equation}
or with any other differential formulation~\cite{conway2011incorporating}.

The performance of an analysis is measured in terms of the variance of the estimate for the parameters of interest, for example in our case the variance of the estimated signal strength $\mu$. We built a differential estimate of the variance using the Fisher information~\cite{fisher1925} of the likelihood in equation~\ref{eq:lh_full} given by
\begin{equation}\label{eq:fisher}
    F_{ij} = \frac{\partial^2}{\partial\theta_i\partial\theta_j}\left(-\log\mathcal{L}(D_\mathrm{H},\bm{\theta})\right).
\end{equation}
Because the maximum likelihood estimator is asymptotically efficient~\cite{cramer1999mathematical,rao1992information}, the variance of the estimates for $\bm{\theta}$ is asymptotically close to

\begin{equation}\label{eq:inv_cov}
V_{ij} = F_{ij} ^ {-1}.
\end{equation}

Assuming the first diagonal element to correspond to the parameter of interest $\mu$, without loss of generality, the loss function to optimize the variance of the estimate for $\mu$ with respect to the free parameters $\bm{\omega}$ of the \gls{NN} function $\bm{f}$ is $V_{00}$.

To be independent of the statistical fluctuations of the observation, the optimization is performed on an Asimov dataset~\cite{cowan2011asymptotic}. This artificial dataset replaces the observation $\bm{d}$ with the nominal expectation for $\bm{s}$ and $\bm{b}$ serving as representative for the median expected outcome of the analysis in presence of the signal plus background hypothesis.

Given the assumption that the dimensionality reduction performed by the \gls{NN} together with the histogram is a sufficient statistic, the optimization can find a function for $\bm{f}$ that gives the best estimate for the parameter of interest $\mu$, similar to a statistical inference performed on the initial high-dimensional dataset $D$ with an unbinned likelihood.

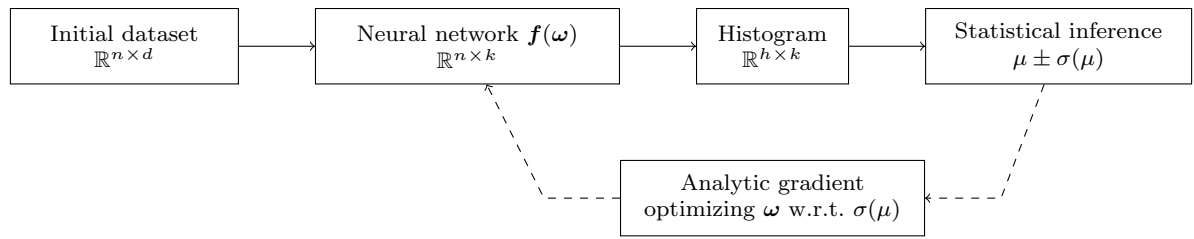
\begin{figure*}
\centering
\begin{tikzpicture}
\node(Initial)[draw,minimum width=3cm,minimum height=1cm,align=center] at (0, 0) {Initial dataset\\$\mathbb{R}^{n\times d}$};
\node(NN)[draw,minimum width=4cm,minimum height=1cm,align=center,right=of Initial] {Neural network $\bm{f}(\bm{\omega})$\\$\mathbb{R}^{n\times k}$};
\node(Hist)[draw,minimum width=2cm,minimum height=1cm,align=center,right=of NN] {Histogram\\$\mathbb{R}^{h\times k}$};
\node(Inference)[draw,minimum width=3.5cm,minimum height=1cm,align=center,right=of Hist] {Statistical inference\\$\mu\pm\sigma(\mu)$};
\node(Grad)[draw,minimum width=4cm,minimum height=1cm,align=center,below=1cm of Hist] {Analytic gradient\\optimizing $\bm{\omega}$ w.r.t. $\sigma(\mu)$};
\draw[->] (Initial) -- (NN);
\draw[->] (NN) -- (Hist);
\draw[->] (Hist) -- (Inference);
\node(Empty1)[shape=coordinate,below=of Inference,right=of Grad] {};
\draw[dashed,->] (Inference) -- (Empty1) -- (Grad);
\node(Empty2)[shape=coordinate,below=of NN,left=of Grad] {};
\draw[dashed,->] (Grad) -- (Empty2) -- (NN);
\end{tikzpicture}
\caption{Graphical overview of the proposed method to optimize the reduction of the dataset used for the statistical inference of the parameters of interest from end to end. The number of observables $d$ in the initial dataset with $n$ observations is reduced to a set of $k$ observables by the neural network function $\bm{f}$ with the free parameters $\bm{\omega}$. The dataset is compressed further by summarizing the $n$ observations using a $k$-dimensional histogram with $h$ bins. Eventually the free parameters $\bm{\omega}$ are optimized with the variance of the parameter of interest $\mu$ as objective, which is made possible by an approximated gradient for the histogram.}
\label{fig:overview}
\end{figure*}

A graphical overview of the proposed method is given in figure~\ref{fig:overview}.

\section{Related work}
\label{sec:Related_work}

The authors of~\cite{deCastro:2018mgh} were first to develop an approach that also optimizes the parameters of an \gls{NN} based on the binned Poisson likelihood of the analysis. They also identify the problem that a histogram has no suitable derivative but follow a different strategy to enable automatic differentiation replacing the counts with means of a softmax function. This approach is also followed by~\cite{neos_2020}.

Likewise Ref.~\cite{elwood2018direct} estimates a count with the sum of the \gls{NN} output values but uses an inclusive estimate of the significance as training objective, which results in an improved analysis objective in a search for new physics.

The strategy to allow a \gls{NN} to find the best compression of the data has also been discussed in~\cite{Charnock:2018ogm}. The authors show that the \gls{NN} is able to learn a summary statistic that is a close approximation of a sufficient statistic, yielding a powerful statistical inference.

A related approach to training the \gls{NN} on the statistical model of the analysis including systematic uncertainties is the explicit decorrelation against the systematic variation. For example, the idea has been discussed on the basis of an adversarial architecture~\cite{louppe2017learning,shimmin2017decorrelated,estrade:hal-01715155}, a penalty term based on distance correlation~\cite{kasieczka2020disco} and an approach penalizing the variation using approximated bin counts~\cite{wunsch2019reducing}. These strategies are not aware of the analysis objective such as the variance of a parameter of interest and therefore the decorrelation is subject to manual optimization. For a large number of nuisances, this optimization procedure is computational expensive and typically unfeasible.

Another approach to optimize the statistical inference is the direct estimation of the likelihood in the input space, which is typically carried out using machine learning techniques. Such methods intend to use the approximated likelihood in the input space for the statistical inference, which avoids the dimensionality reduction that is optimized with the proposed method.  The technical difficulties to carry out these methods are discussed in~\cite{cranmer2019frontier,cranmer2015approximating}.

\section{Application to a simple example based on pseudo-experiments}
\label{sec:Application_to_a_simple_example}

\begin{figure}
\centering
\includegraphics[width=0.8\linewidth]{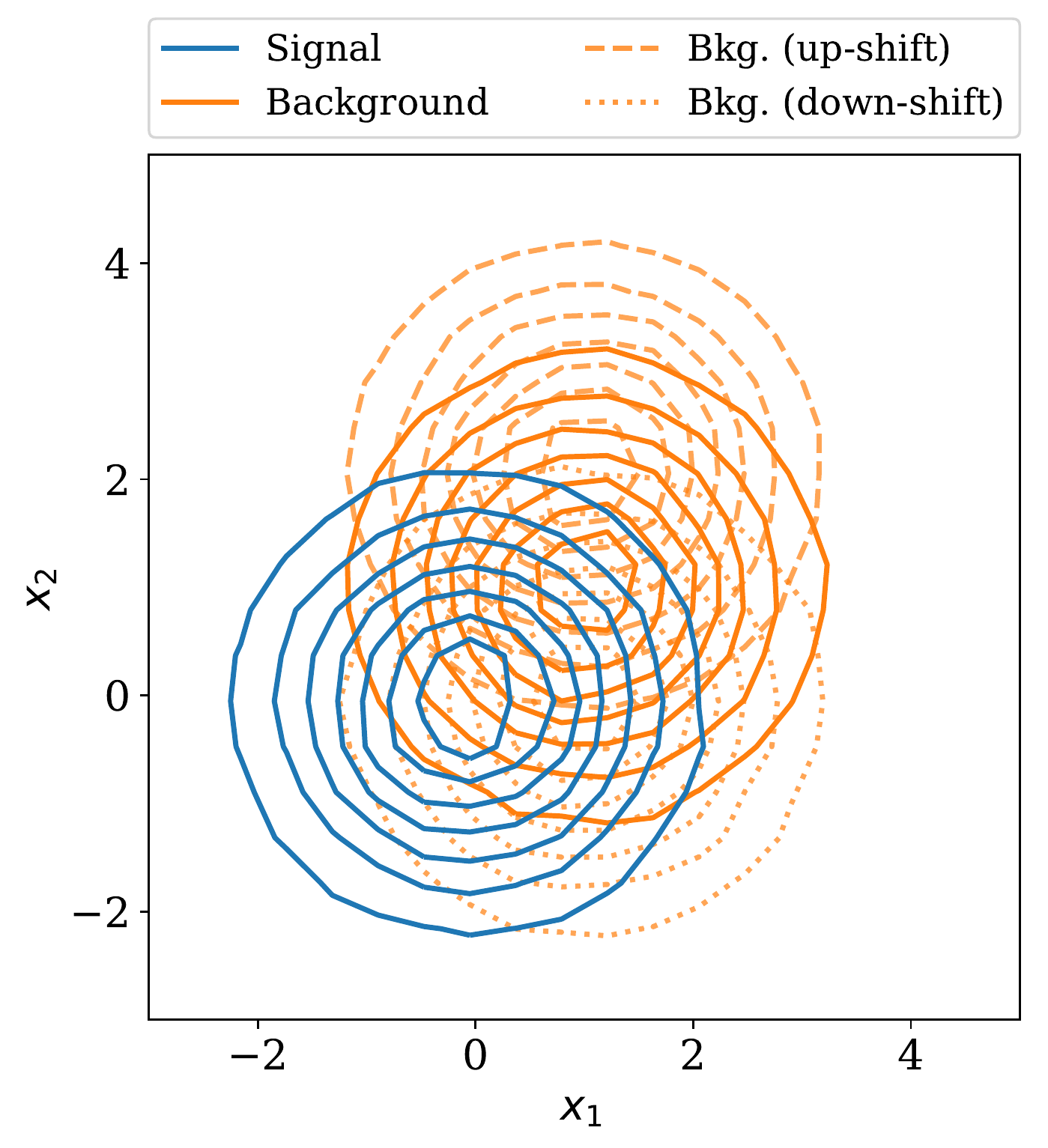}
\caption[]{Distribution of the signal and background components in the input space modelled by multivariate Gaussian distributions centered around $\begin{psmallmatrix} 0 & 0\end{psmallmatrix}$ and $\begin{psmallmatrix} 1 & 1\end{psmallmatrix}$ with the covariance matrix $\begin{psmallmatrix} 1 & 0 \\ 0 & 1 \end{psmallmatrix}$. We introduce a systematic variation that shifts the mean of the background component along $x_2$.}
\label{fig:toy_data}
\end{figure}

A simple example based on pseudo-experiments and a known likelihood in the input space $\mathbb{R}^{n\times d}$ is used to illustrate our approach. The distributions of the signal and background components in the input space are shown in figure~\ref{fig:toy_data}. We assume a systematic uncertainty on the mean of the background process modelled  by the shifts $x_2 \pm 1$, representing the systematic variations in equation~\ref{eq:nuisance_asymmetric}.

The \gls{NN} architecture is a fully-connected feed-forward network with $100$ nodes in one hidden layer. The initialization follows the Glorot algorithm~\cite{glorot2010understanding} and the activation function is a rectified linear unit~\cite{glorot2011deep}. The output layer has a single node with a sigmoid activation function.

We use eight bins for the histogram of the \gls{NN} output and compute the variance of the estimate for the parameter of interest $\mu$ denoted by $V_{00}$. The operations are implemented using TensorFlow as computational graph library~\cite{abadi2016tensorflow,tensorflow_probability} and we use the provided automatic differentiation and the Adam algorithm~\cite{kingma2014adam} to optimize the free parameters $\bm{\omega}$ with the objective to minimize $V_{00}$. The systematic variations $\bm{\Delta}$ can be implemented with reweighting techniques using statistical weights or duplicates of the nominal dataset with the simulated variations, whereas we chose the latter solution. Each gradient step is performed on the full dataset with \SI{e5}{} simulated events for each process. The dataset is split in half for training and validation, and all results are computed from a statistically independent dataset of the same size as the original one. The training is stopped if the loss has not improved for $\SI{100}{}$ gradient steps eventually using the model with the smallest loss on the validation dataset for further analysis. We found that the convergence is more stable if the model is first optimized only on the statistical part of the likelihood shown in equation~\ref{eq:lh_stats} and therefore apply this pretraining for $30$ gradient steps. We apply statistical weights to scale the expectation of signal and background to $\SI{50}{}$ and $\SI{e3}{}$, respectively.

\begin{figure}
\centering
\includegraphics[width=0.8\linewidth]{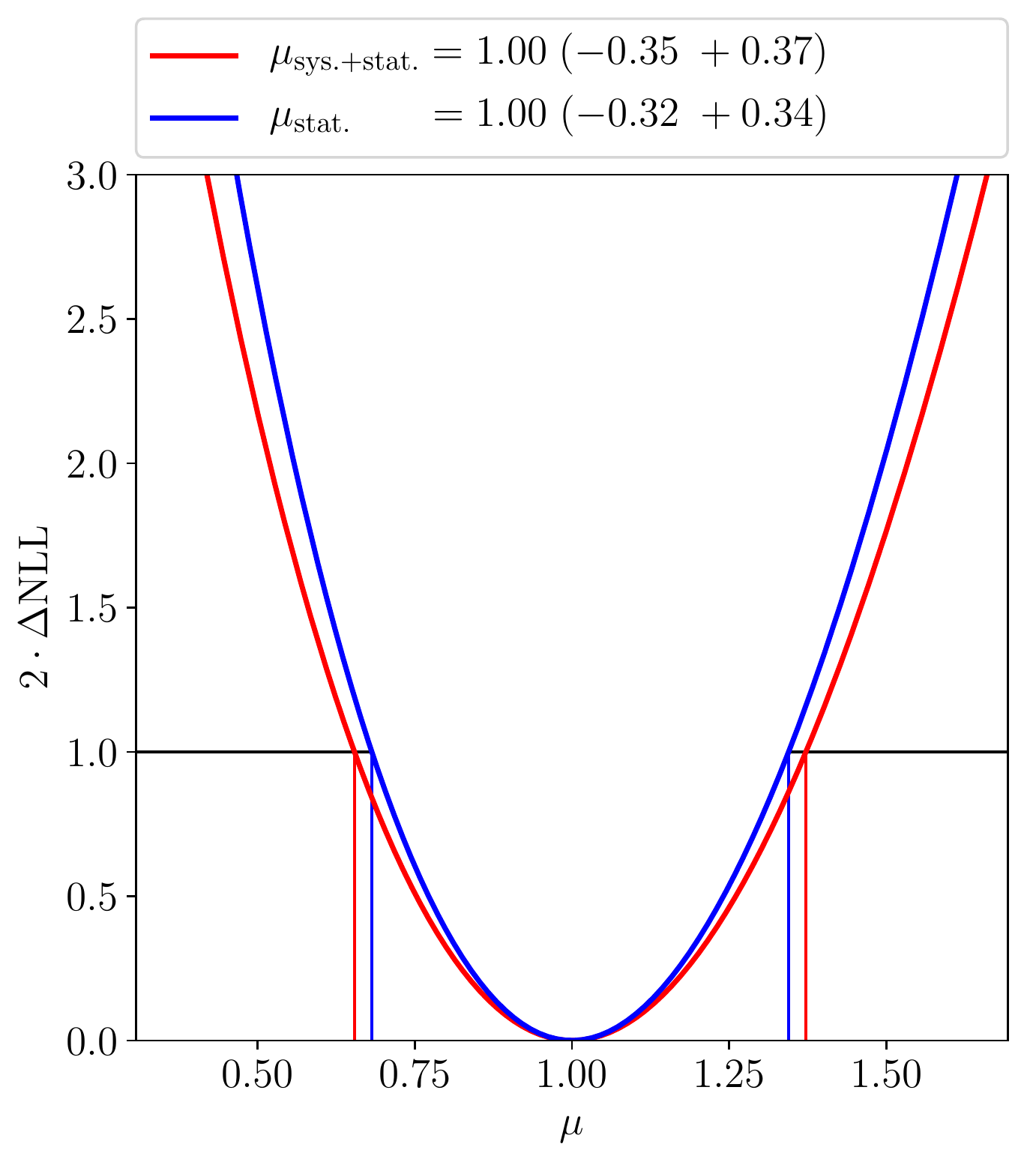}
\caption{Profile of the likelihood with (blue line) only the statistical uncertainty and (red line) the systematic uncertainty in addition for the likelihood defined in the two-dimensional input space spanned by $x_1$ and $x_2$ as given in Fig.~\ref{fig:toy_data}}
\label{fig:toy_opt}
\end{figure}

The best possible expected result in terms of the variance of the estimate for $\mu$ is given by a fit of the unbinned statistical model without dimensionality reduction. Alternatively, we can get an asymptotically close result by using a binned likelihood with sufficiently large number of bins in the two-dimensional input space. The latter approach with $20\times 20$ equidistant bins in the range shown in figure~\ref{fig:toy_data} results in the profile shown in figure~\ref{fig:toy_opt} with $\mu=1.0^{+0.37}_{-0.35}$. The best-fit value of $\mu$ is always at $1.0$ because of the used Asimov dataset. Further, we find the uncertainty of $\mu$ in all fits by profiling the likelihood~\cite{james2006statistical} rather than using the approximation by the covariance matrix in equation~\ref{eq:inv_cov}. We obtain all results in this paper with validated statistical tools, RooFit and RooStats~\cite{root,RooStats,roofit}, such as used by most publications analyzing data of the \gls{LHC} experiments.

\begin{figure*}[hp]
\centering
\includegraphics[width=0.31\linewidth]{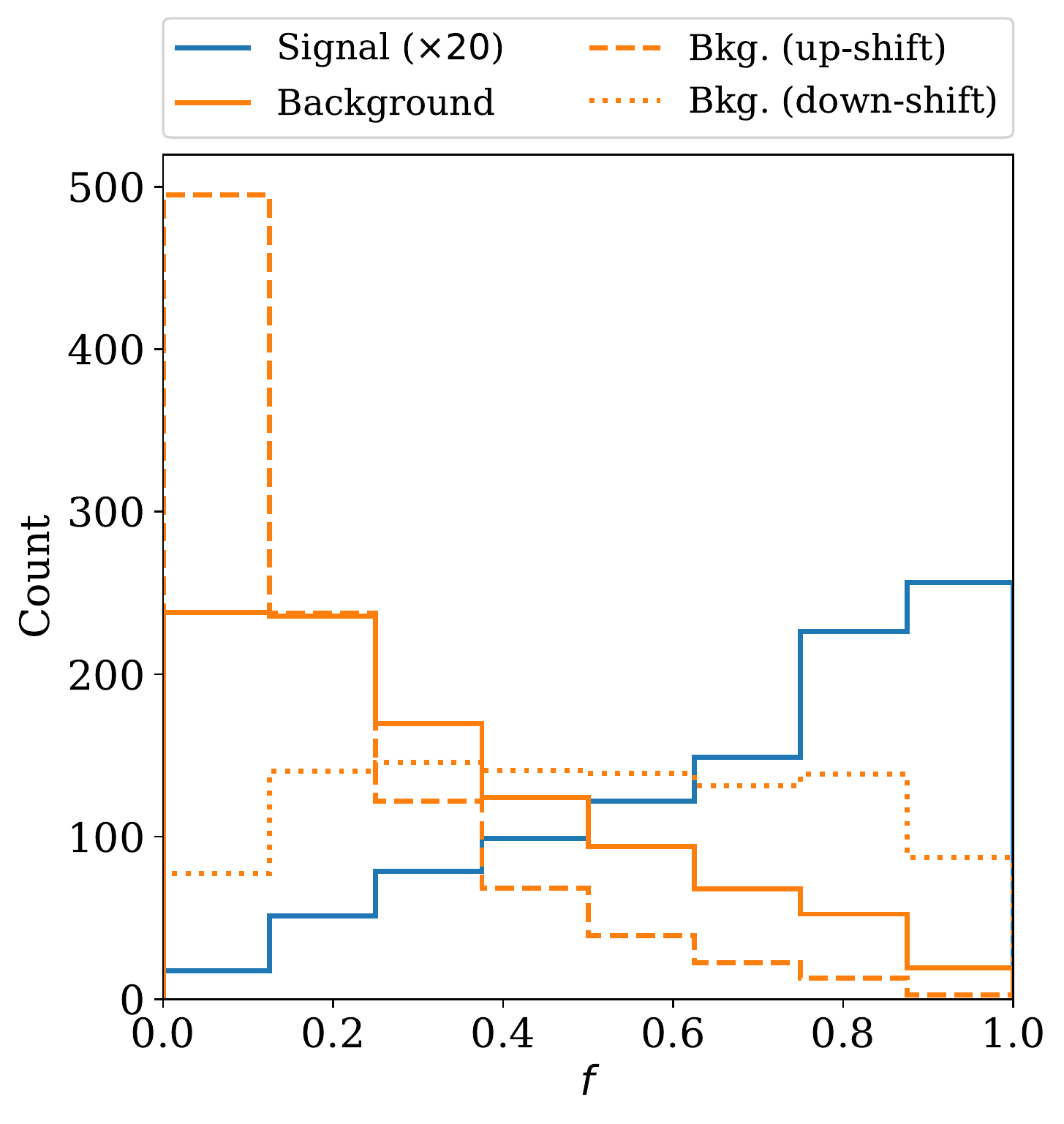}%
\includegraphics[width=0.35\linewidth]{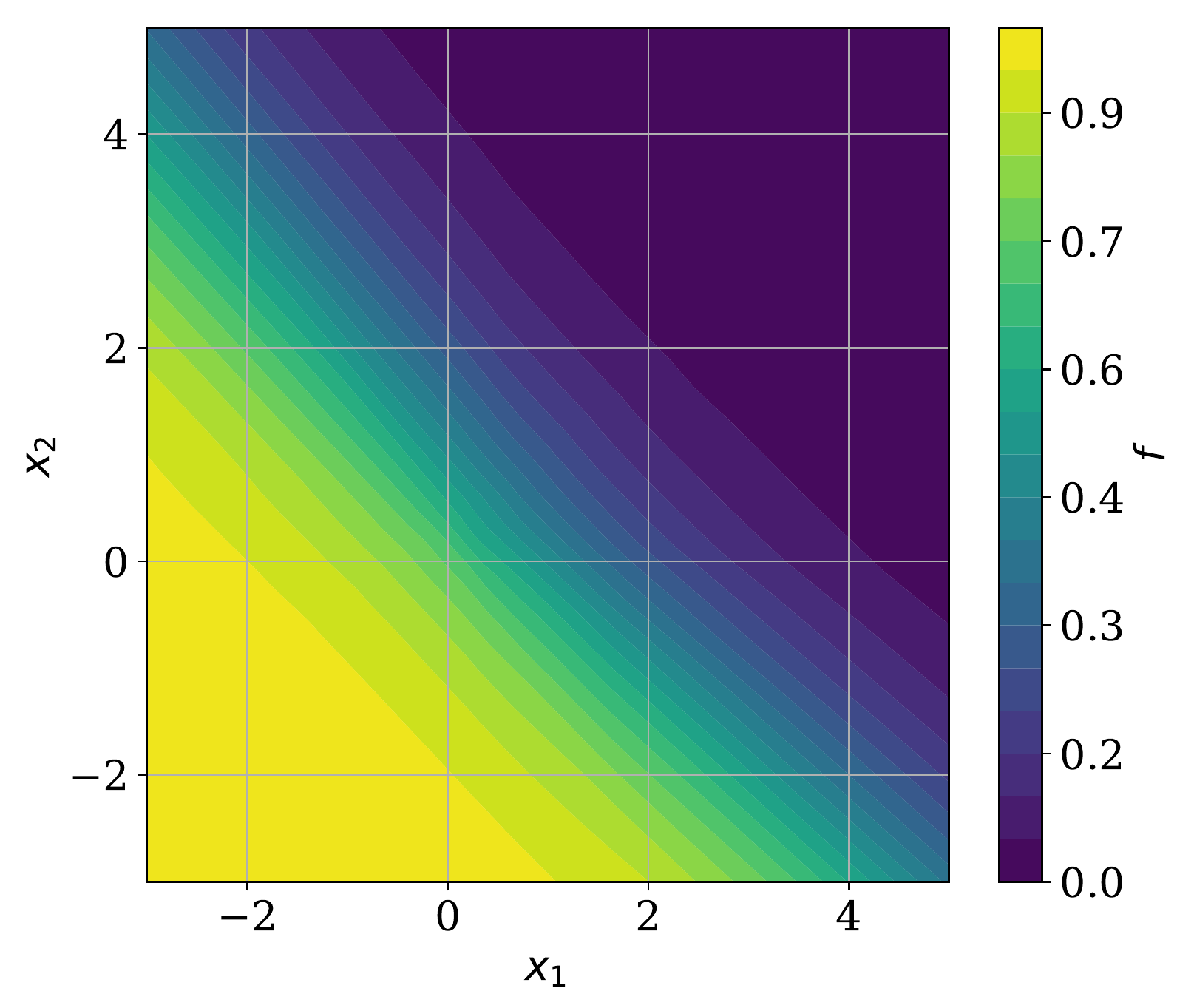}%
\includegraphics[width=0.30\linewidth]{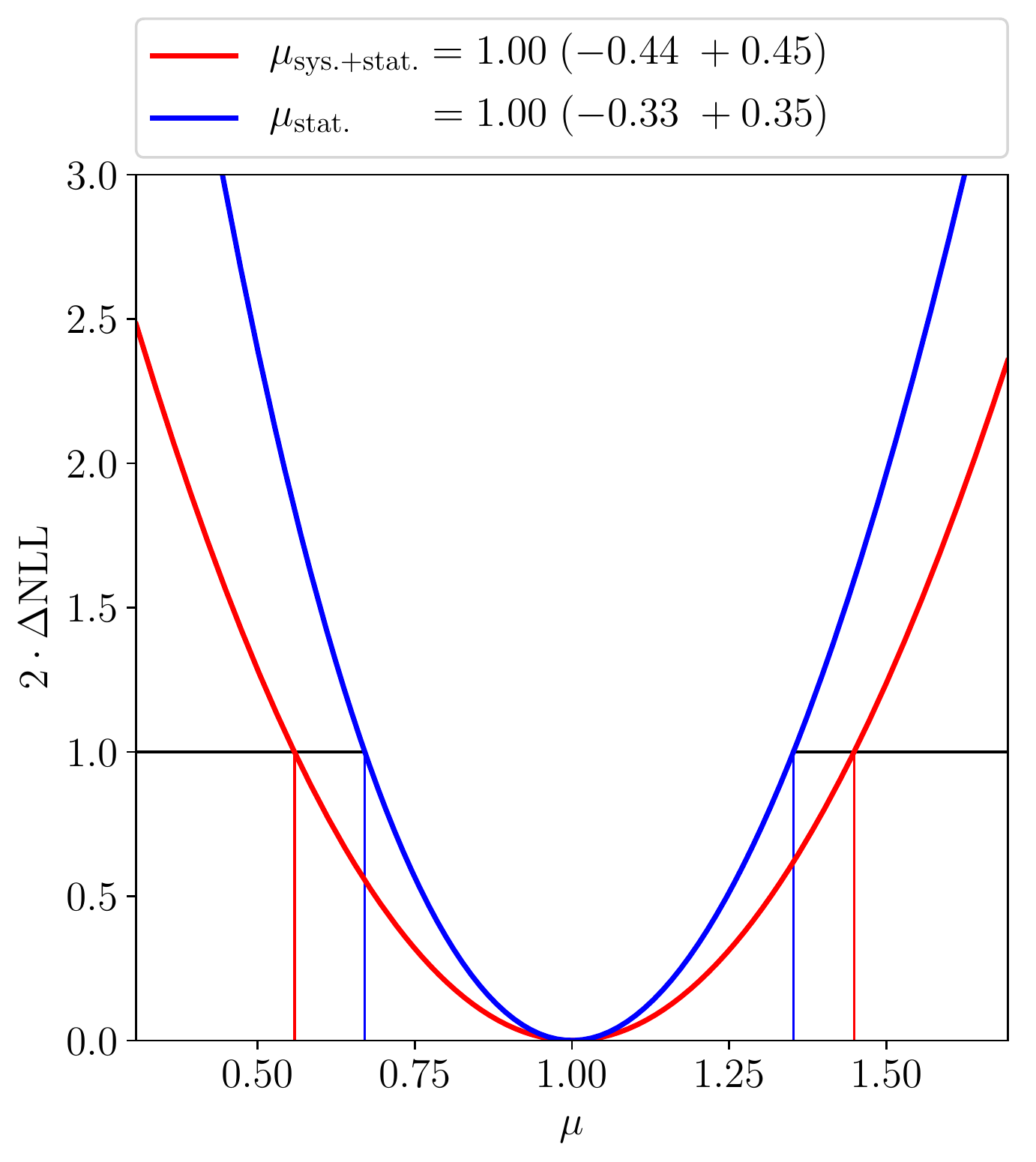}
\caption{Distributions of the \gls{NN} output for the simple example consisting of signal, background, and systematic variation in the (left) input space spanned by $x_{1}$ and $x_{2}$ and (middle) value space, if the \gls{NN} is trained on the classification of the two processes using the cross entropy loss. The likelihood profiles taking (red line) only the statistical uncertainty and (blue line) the statistical and systematic uncertainty into account for the final statistical inference of $\mu$ are shown on the right.}
\label{fig:toy_ceonly}

\includegraphics[width=0.31\linewidth]{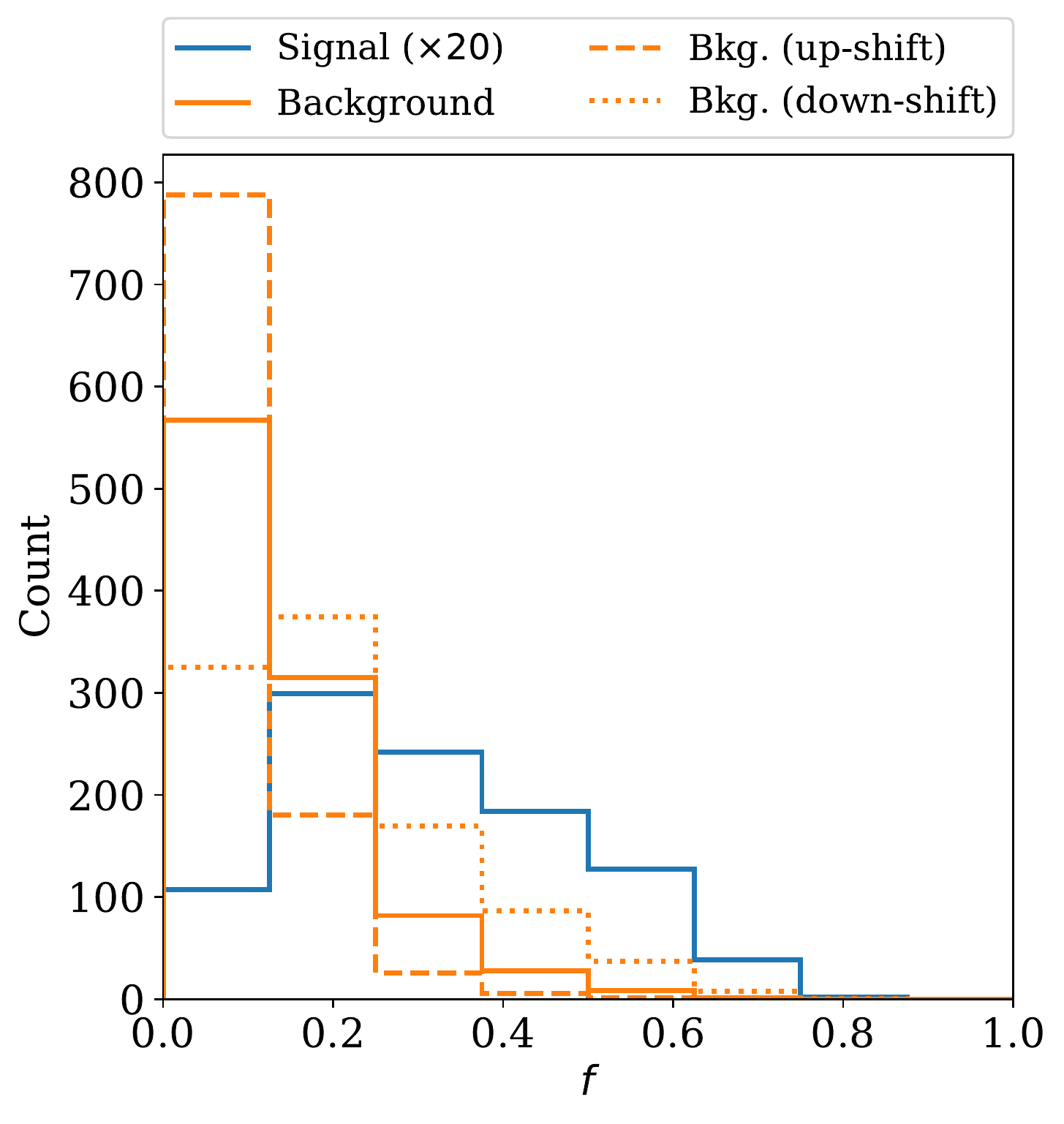}%
\includegraphics[width=0.35\linewidth]{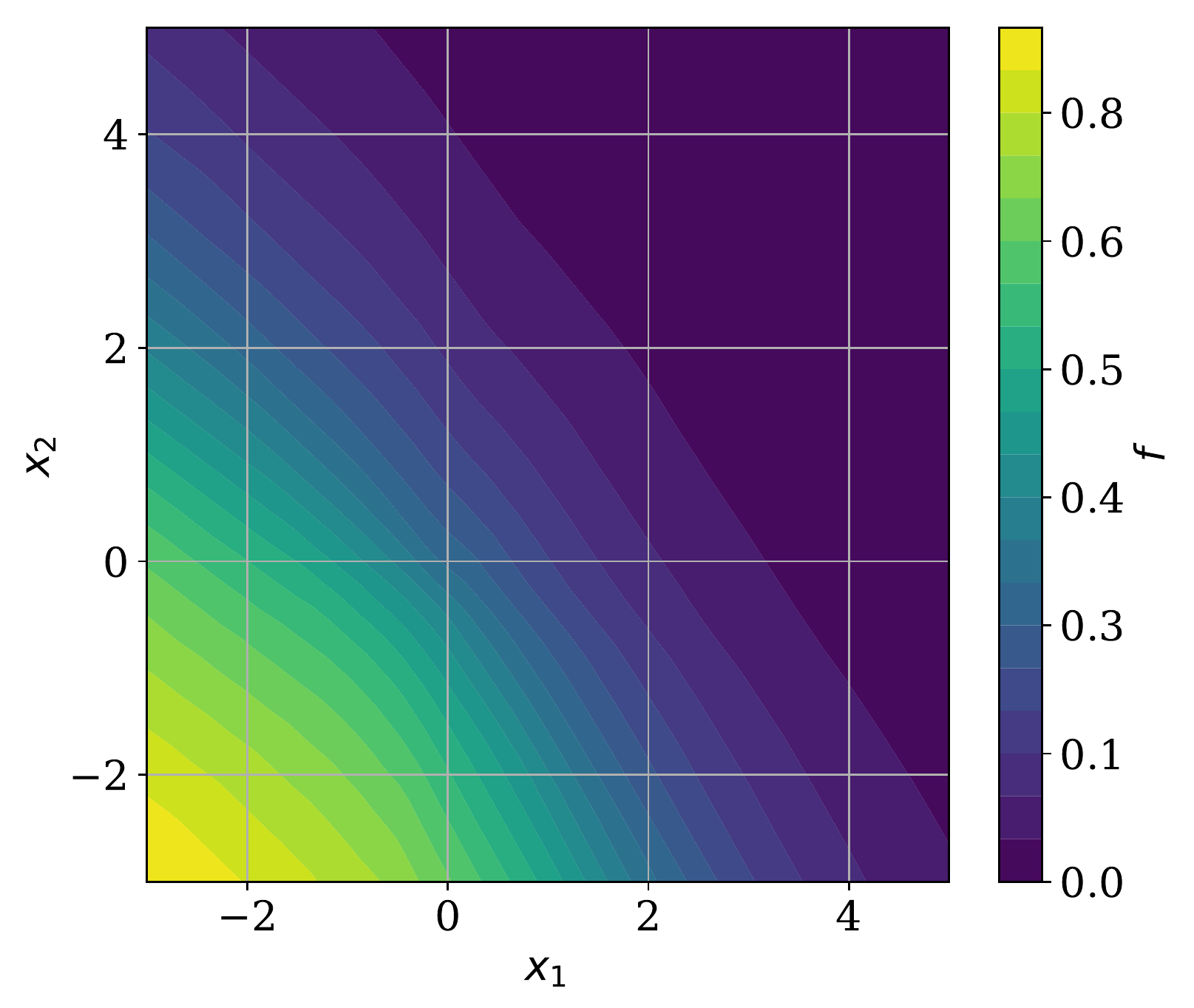}%
\includegraphics[width=0.30\linewidth]{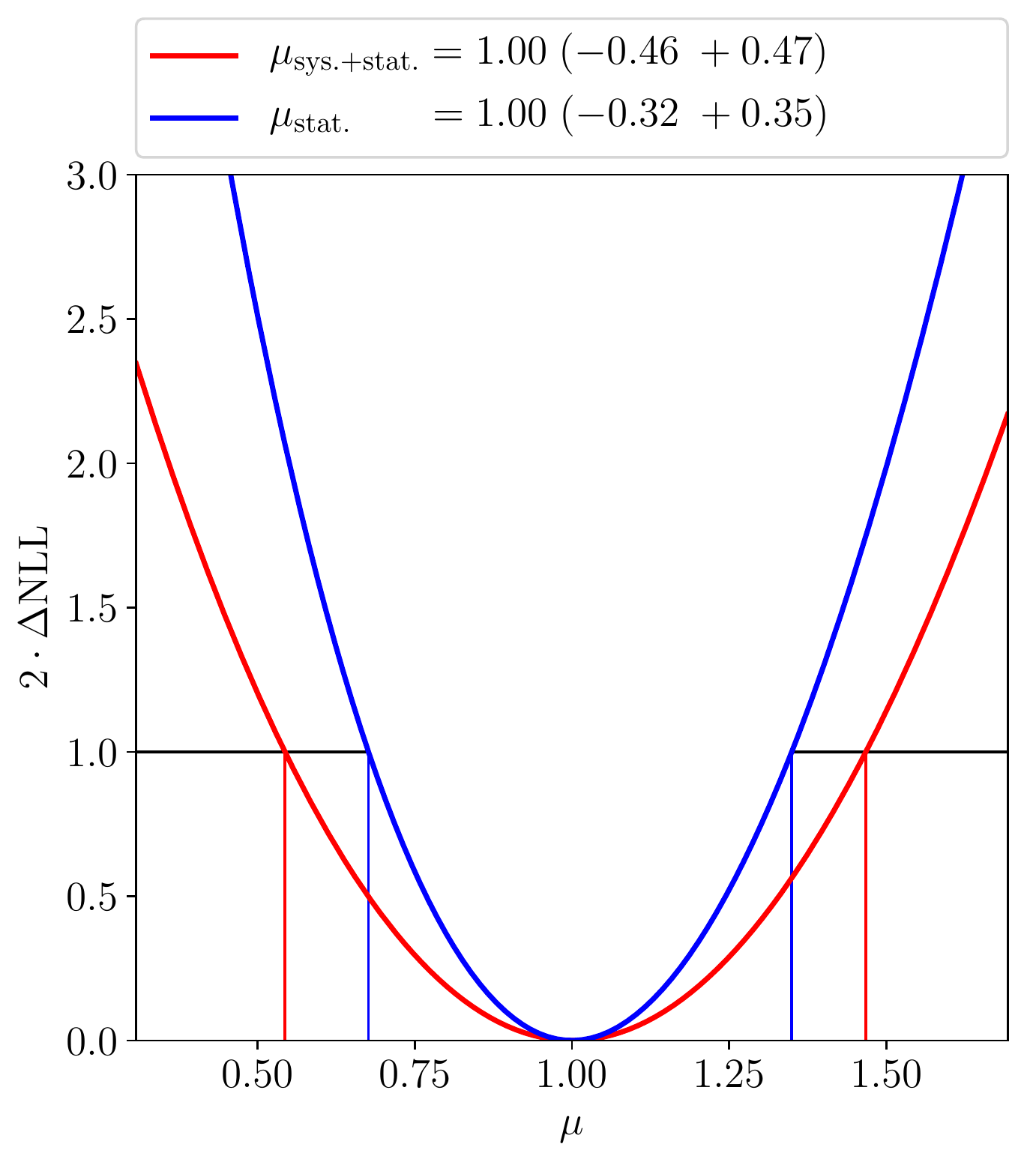}
\caption{Distributions of the \gls{NN} output for the simple example consisting of signal, background, and systematic variation in the (left) input space spanned by $x_{1}$ and $x_{2}$ and (middle) value space, if the NN is trained on the variance of the signal strength $V_{00}$ defined by the likelihood without the description of the systematic uncertainty. The likelihood profiles taking (red line) only the statistical uncertainty and (blue line) the statistical and systematic uncertainty into account for the final statistical inference of $\mu$ are shown on the right.}
\label{fig:toy_statsonly}

\includegraphics[width=0.31\linewidth]{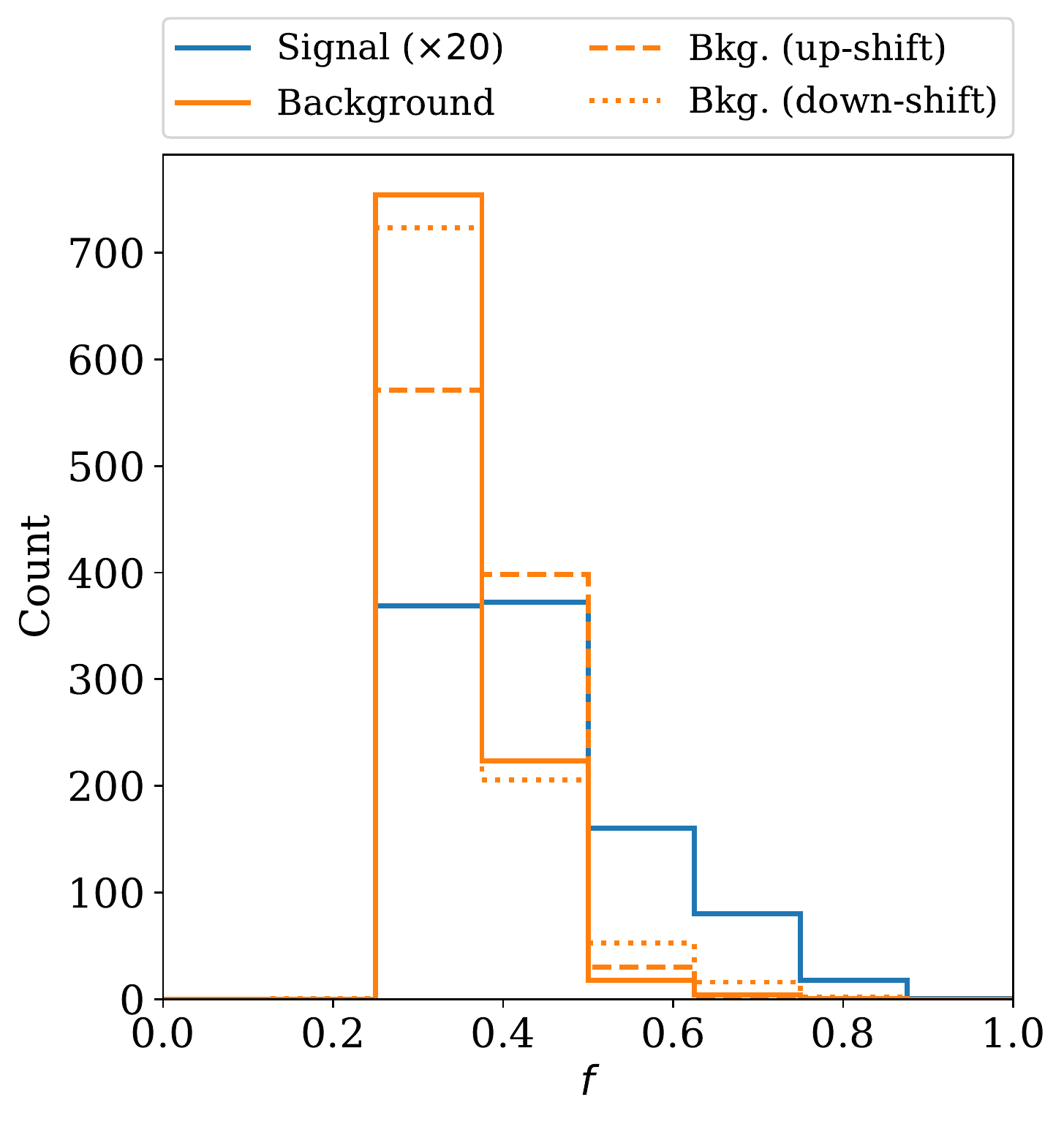}%
\includegraphics[width=0.35\linewidth]{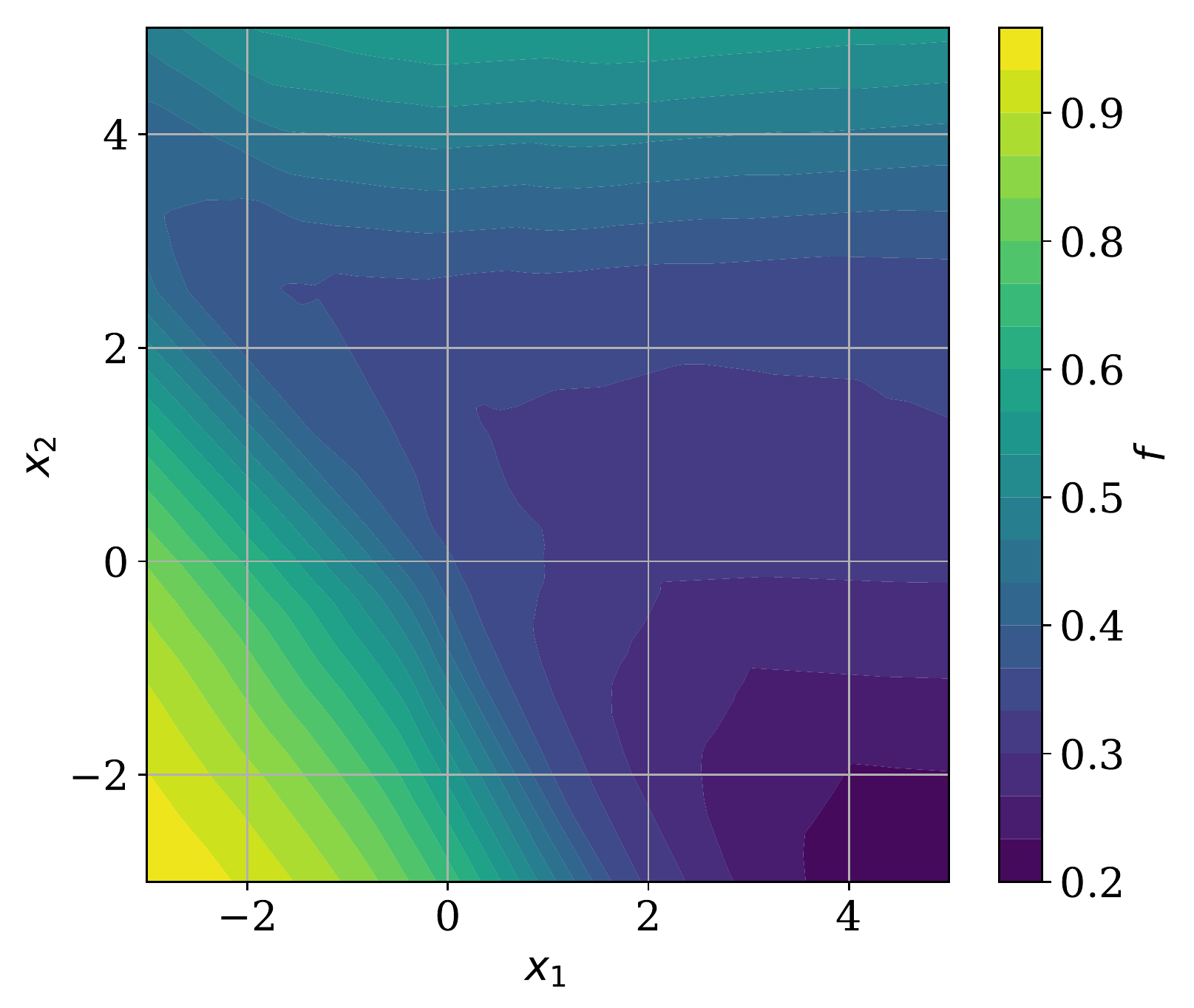}%
\includegraphics[width=0.30\linewidth]{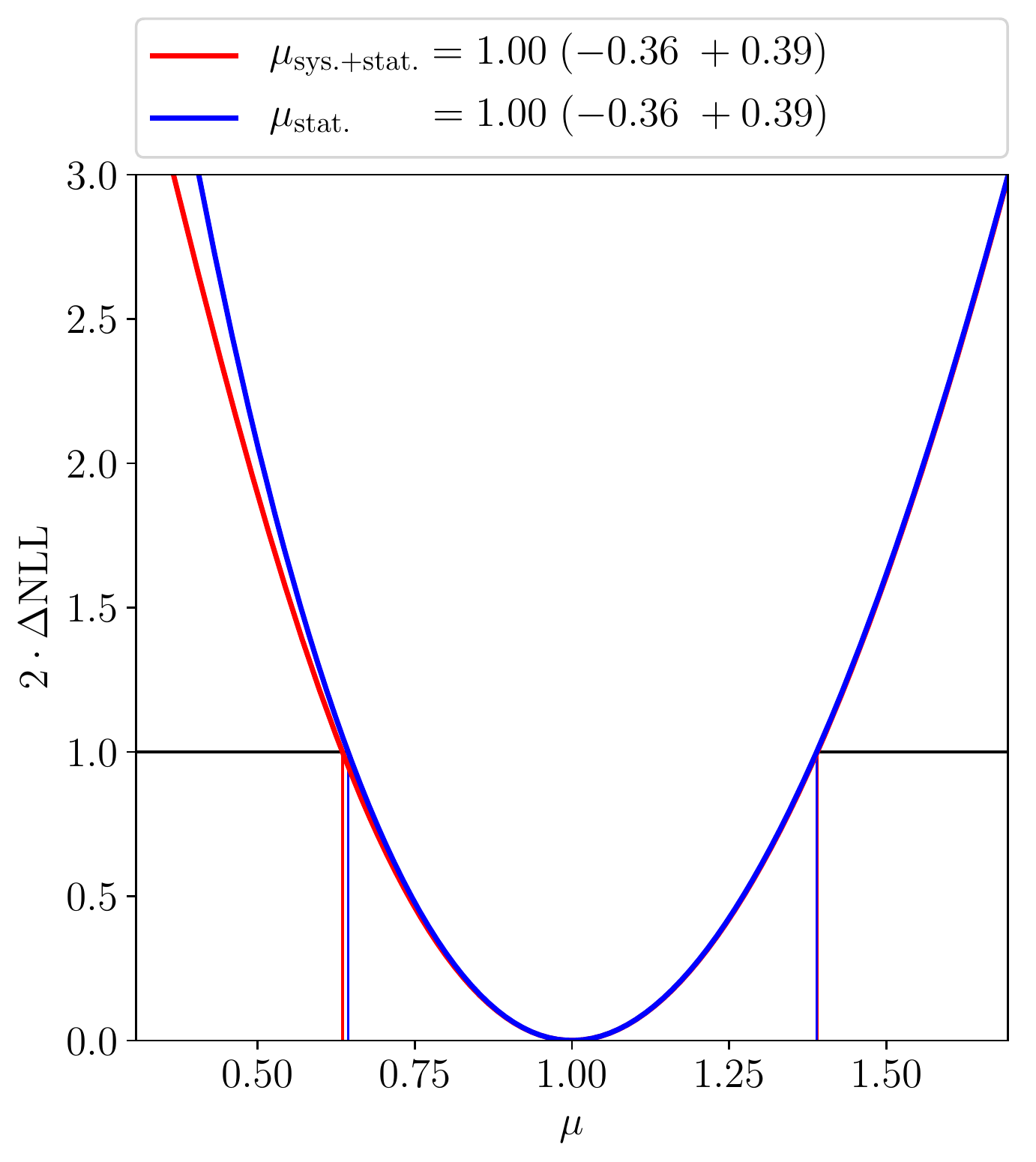}
\caption{Distributions of the \gls{NN} output for the simple example consisting of signal, background, and systematic variation in the (left) input space spanned by $x_{1}$ and $x_{2}$ and (middle) value space, if the NN is trained on the variance of the signal strength $V_{00}$ defined by the likelihood including the systematic uncertainty. The likelihood profiles taking (red line) only the statistical uncertainty and (blue line) the statistical and systematic uncertainty into account for the final statistical inference of $\mu$ are shown on the right.}
\label{fig:toy_fullnll}
\end{figure*}

The first comparison to this best-possible result is done by training the \gls{NN} not on the variance of the estimate for $\mu$, $V_{00}$, but on the cross entropy loss with signal and background weighted to the same expectation. This approach has been used in multiple analyses in high-energy particle physics~\cite{tthbb,smhtt}. The \gls{NN} function $\bm{f}$ is a sufficient statistic - and therefore optimal - if no systematic uncertainties have to be considered for the statistical inference such as the likelihood in equation~\ref{eq:lh_stats}~\cite{deCastro:2018mgh}. The resulting function $\bm{f}$ is shown in the input space and by the distribution of the output in figure~\ref{fig:toy_ceonly}. The \gls{NN} learns to project the two-dimensional space spanned by $x_1$ and $x_2$ on the diagonal, which is trivially the optimal dimensionality reduction in this simple example. If we apply the statistical model including the systematic uncertainty on the histograms in figure~\ref{fig:toy_ceonly}, the parameter of interest is fitted as $\mu=1.0^{+0.45}_{-0.44}$ with an uncertainty worse by $\SI{19}{\percent}$ than the best possible result obtained above, measured with the width of the total error bars.

As a consistency check for our new strategy described in section~\ref{sec:Method}, we train the \gls{NN} on the variance of the estimate for $\mu$ given by $V_{00}$ in equation~\ref{eq:inv_cov} but without adding the nuisance parameter $\eta$ modelling the systematic uncertainty. The resulting \gls{NN} function $\bm{f}$ in the input space, the distribution of the outputs and the profile of the likelihood are shown in figure~\ref{fig:toy_statsonly}. As expected, the plane of the function $\bm{f}$ in the input space is qualitatively similar, resulting with $\mu=1.0^{+0.47}_{-0.46}$ in a comparable performance than the training on the cross entropy loss. It should be noted that the systematic uncertainty has been included again for the statistical inference.

When adding the nuisance parameter $\eta$ to the likelihood, the training of the \gls{NN} results in the function $\bm{f}$ shown in figure~\ref{fig:toy_fullnll}. The uncertainty of the parameter of interest is with the fit result $\mu=1.0^{+0.39}_{-0.36}$ considerably decreased and lowers the residual difference to the optimal result from $\SI{19}{\percent}$ to $\SI{4}{\percent}$. The function $\bm{f}$ in the input space in figure~\ref{fig:toy_fullnll} shows that the training identified successfully the signal-enriched region with less contribution of the systematic uncertainty resulting in counts in the histogram yielding high signal statistics with a small uncertainty from the variation of the background process. Figure~\ref{fig:toy_fullnll} shows also that the \gls{NN} function is decorrelated against the systematic uncertainty because the profile of the likelihood changes only little if we remove the systematic uncertainty from the statistical model. The proposed method shares this feature with other approaches for decorrelation of the \gls{NN} function such as discussed in section~\ref{sec:Related_work}. The difference is that the strength of the decorrelation is not a hyperparameter but controlled by the higher objective $V_{00}$, which enables us to find directly the best trade-off between statistical and systematic uncertainty contributing to the estimate of $\mu$. The correlation of the parameter of interest $\mu$ to the parameter $\eta$ controlling the systematic variation is reduced from $\SI{64}{\percent}$ for the training on the cross entropy loss to $\SI{13}{\percent}$ for the training on the variance of the parameter of interest $V_{00}$.

\section{Application to a more complex analysis task typical for high-energy particle physics}
\label{sec:Application_to_a_more_complex}

\begin{figure*}[p]
\begin{minipage}[c]{0.4\textwidth}
\caption{Distribution of the missing transverse energy (\texttt{PRI\_met}) for the (left) signal and (right) background process}
\label{fig:higgs_met}
\end{minipage}\hfill
\begin{minipage}[c]{0.6\textwidth}
\includegraphics[width=0.5\linewidth]{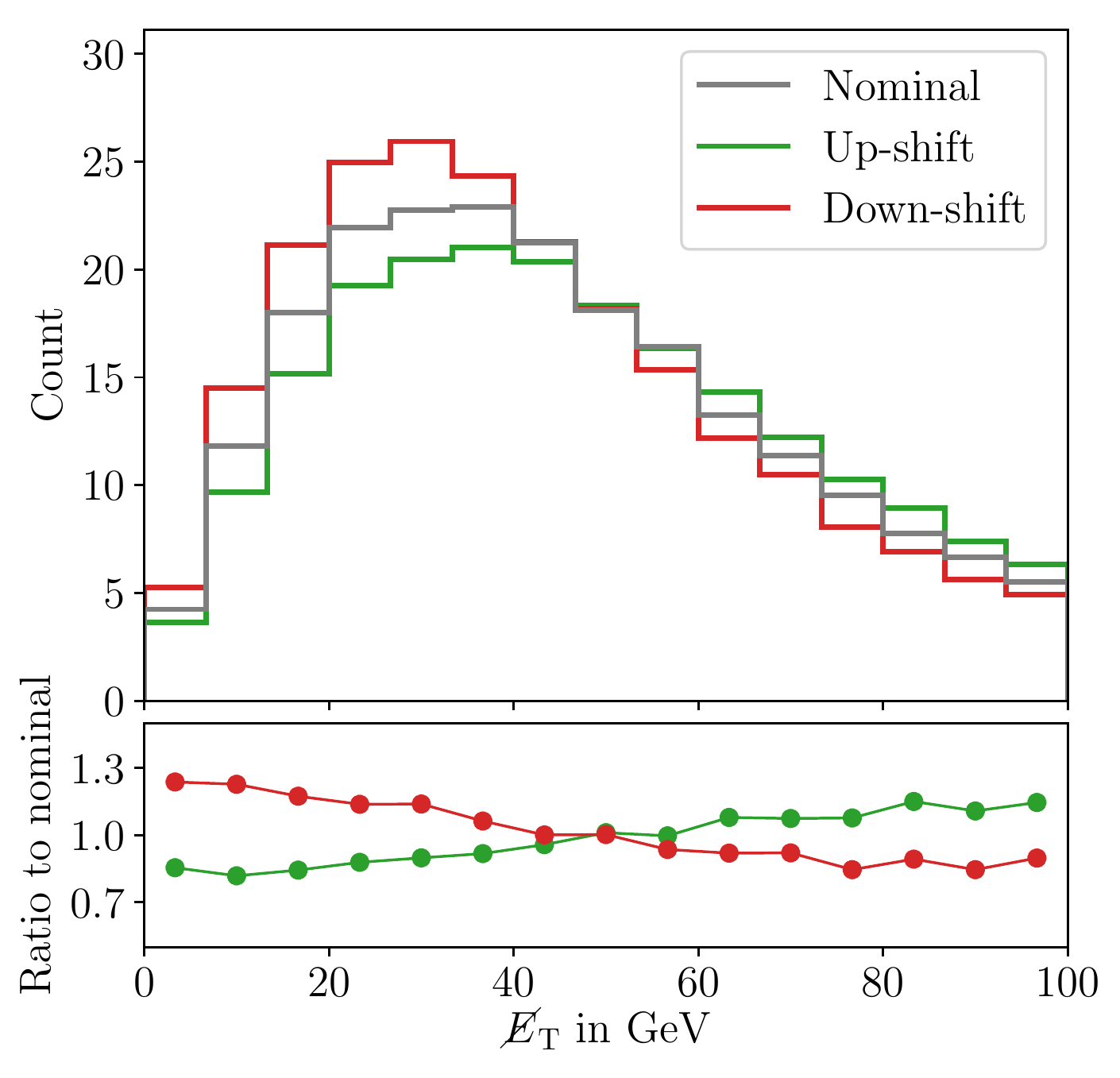}%
\includegraphics[width=0.5\linewidth]{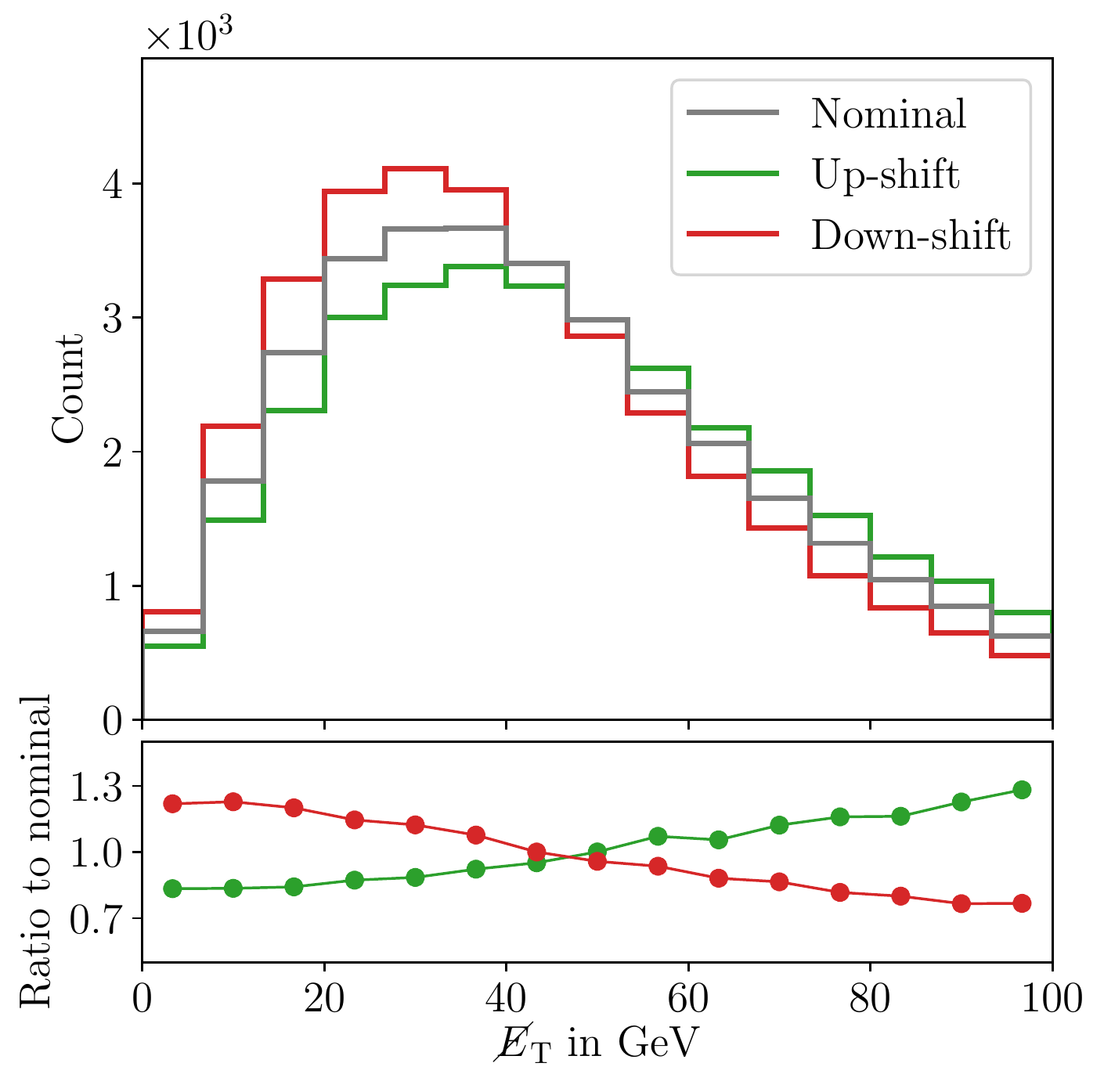}
\end{minipage}
\end{figure*}

\begin{figure*}[p]
\begin{minipage}[c]{0.4\textwidth}
\caption{Distribution of the visible mass of the di-tau system (\texttt{DER\_mass\_vis}) for the (left) signal and (right) background process}
\label{fig:higgs_mass_vis}
\end{minipage}\hfill
\begin{minipage}[c]{0.6\textwidth}
\includegraphics[width=0.5\linewidth]{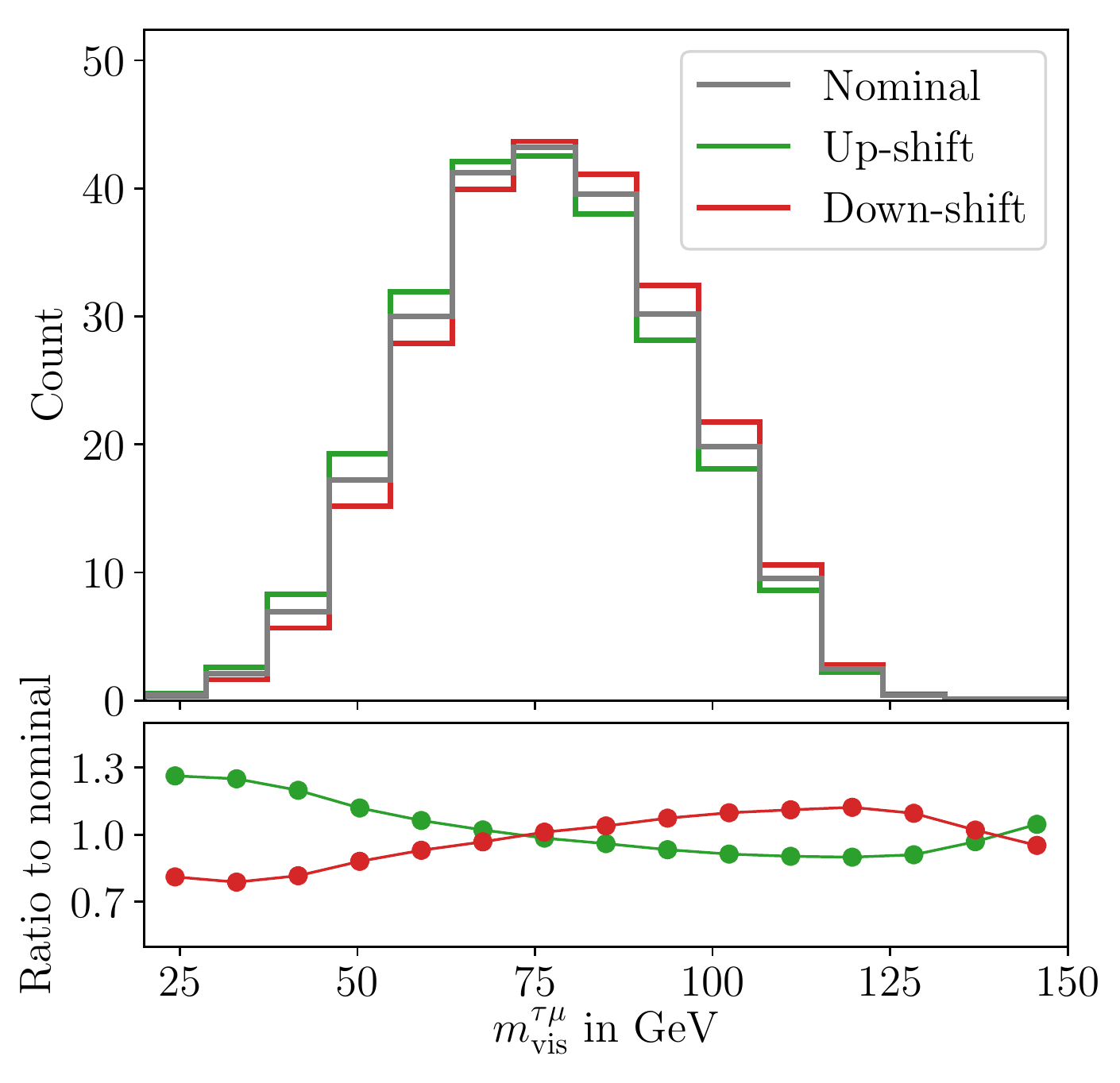}%
\includegraphics[width=0.5\linewidth]{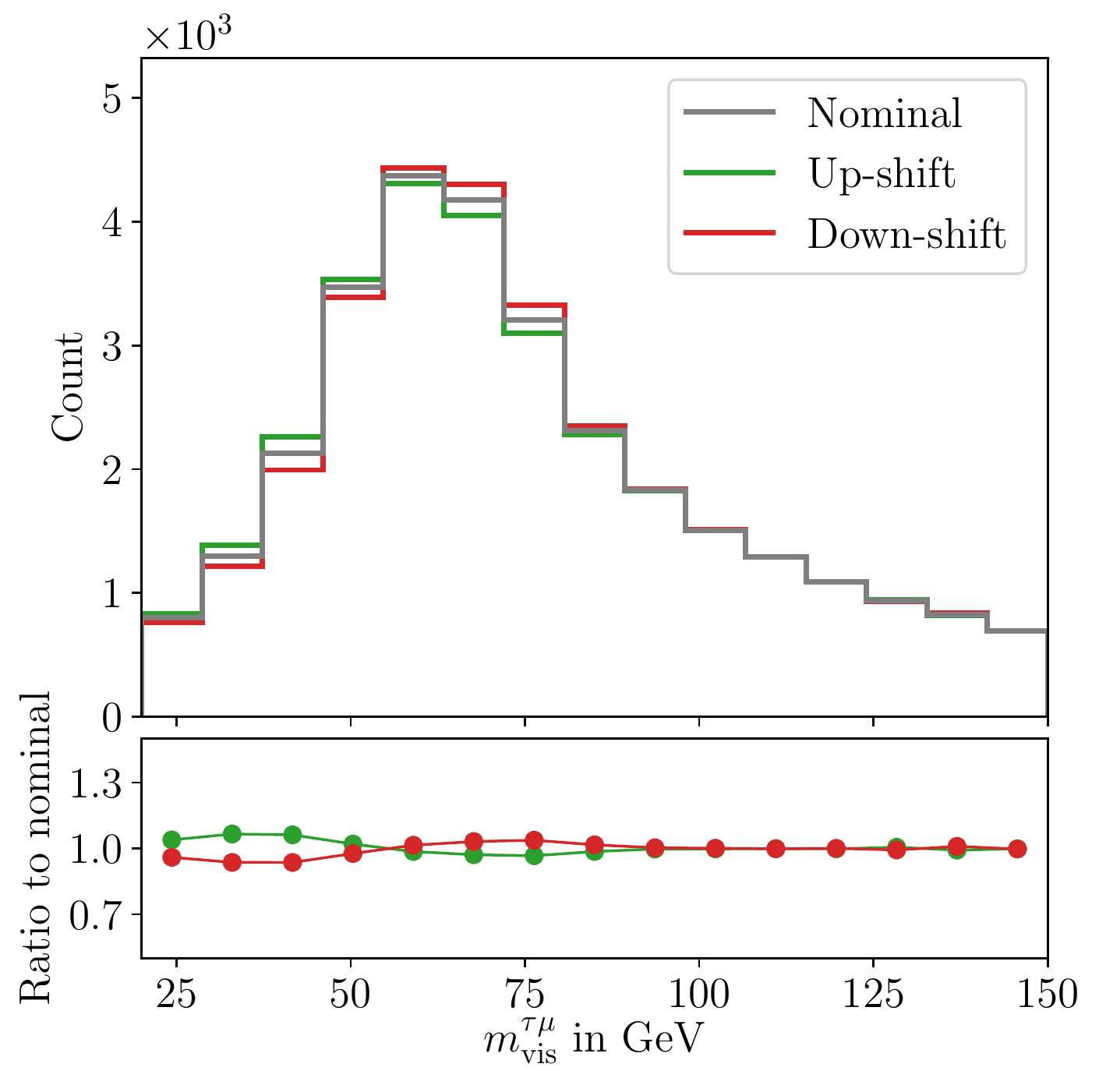}
\end{minipage}
\end{figure*}

\begin{figure*}[p]
\begin{minipage}[c]{0.4\textwidth}
\caption{Distribution of the transverse momentum built from the vector sum of the hadronic tau, the muon and the missing transverse momentum (\texttt{DER\_pt\_h}), used as an estimate of the transverse momentum of the reconstructed Higgs boson candidate, for the (left) signal and (right) background process}
\label{fig:higgs_pt}
\end{minipage}\hfill
\begin{minipage}[c]{0.6\textwidth}
\includegraphics[width=0.5\linewidth]{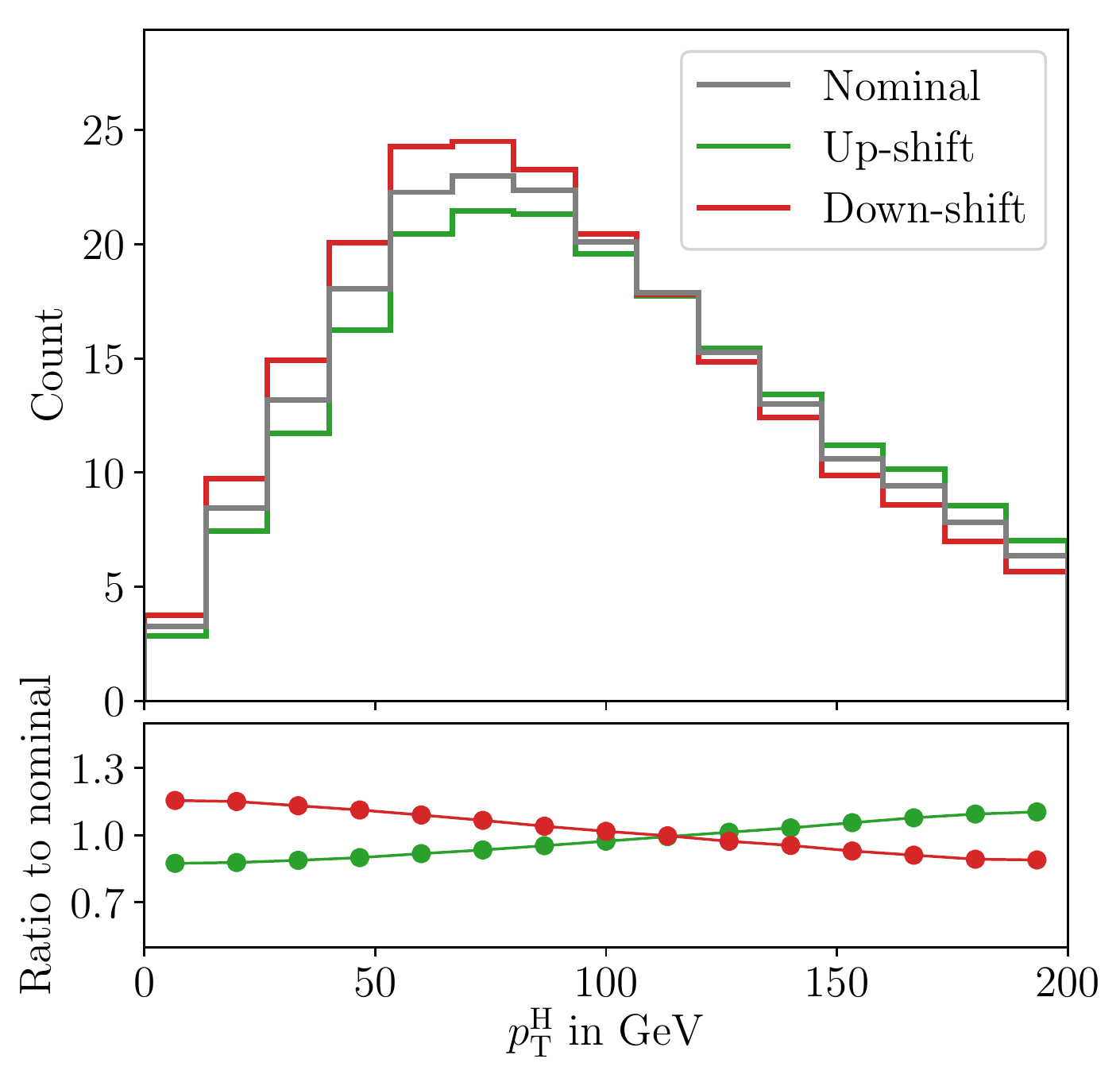}%
\includegraphics[width=0.5\linewidth]{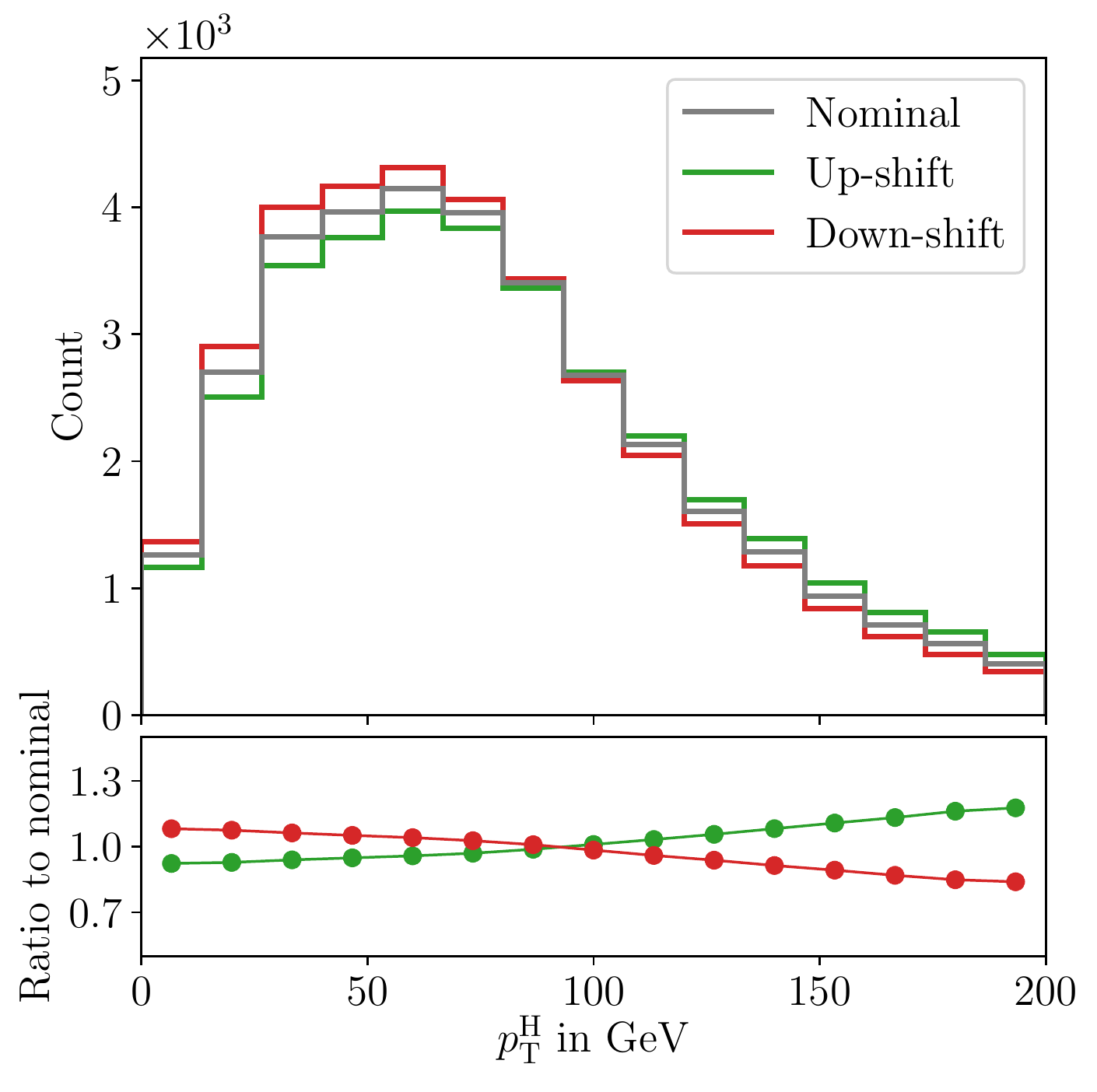}
\end{minipage}
\end{figure*}

In this section, we apply the proposed method to a problem typical for data analysis in high-energy particle physics at the~\gls{LHC}. We use a subset of the dataset published for the Higgs boson machine learning challenge~\cite{adam2014learning,kaggle_higgs_data} extended by a systematic variation. The goal of the challenge is to achieve the best possible significance for the signal process representing Higgs boson decays to two tau leptons overlaid by the background simulated as a mixture of different physical processes~\cite{adam2014learning}. We pick from the dataset four variables, namely \texttt{PRI\_met}, \texttt{DER\_mass\_vis}, \texttt{DER\_pt\_h} and\linebreak\texttt{DER\_deltaeta\_jet\_jet} and select only events, which have all of these features defined. In addition to the event weights provided with the dataset, we scale the signal expectation with a factor of two. The final dataset has 244.0 and 35140.1 (106505 and 131480) weighted (unweighted) events for the signal and background process, respectively. The systematic uncertainty in the dataset is assumed as a $\SI{10}{\percent}$ uncertainty on the missing transverse energy implemented with the transformation $\texttt{PRI\_met} \cdot (1.0 \pm 0.1)$ and propagated to the other variables using reweighting. The distributions of the variables including the systematic variations are shown in figures~\ref{fig:higgs_met} to \ref{fig:higgs_pt}. The \gls{NN} is trained only on three of the four variables, excluding the missing transverse energy. The systematic variations propagated to the remaining variables are thus correlated via a hidden variable, representing a more complex scenario than the simple example in section~\ref{sec:Application_to_a_simple_example}. We split the dataset using one third for training and validation of the \gls{NN}, and two thirds for the results presented in this paper. The \gls{NN} architecture and the training procedure are the same as implemented for the simple example in section~\ref{sec:Application_to_a_simple_example} with the difference that we apply a standardization of the input ranges following the rule $(x - \overline{x})/\sigma(x)$ with the mean $\overline{x}$ and standard variation $\sigma(x)$ of the input $x$.

\begin{figure*}[p]
\begin{minipage}[c]{0.4\textwidth}
\caption{Distribution of the absolute difference in the pseudorapidity of the two leading jets (\texttt{DER\_deltaeta\_jet\_jet}) for the (left) signal and (right) background process}
\label{fig:higgs_deltaeta_jet_jet}
\end{minipage}\hfill
\begin{minipage}[c]{0.6\textwidth}
\includegraphics[width=0.5\linewidth]{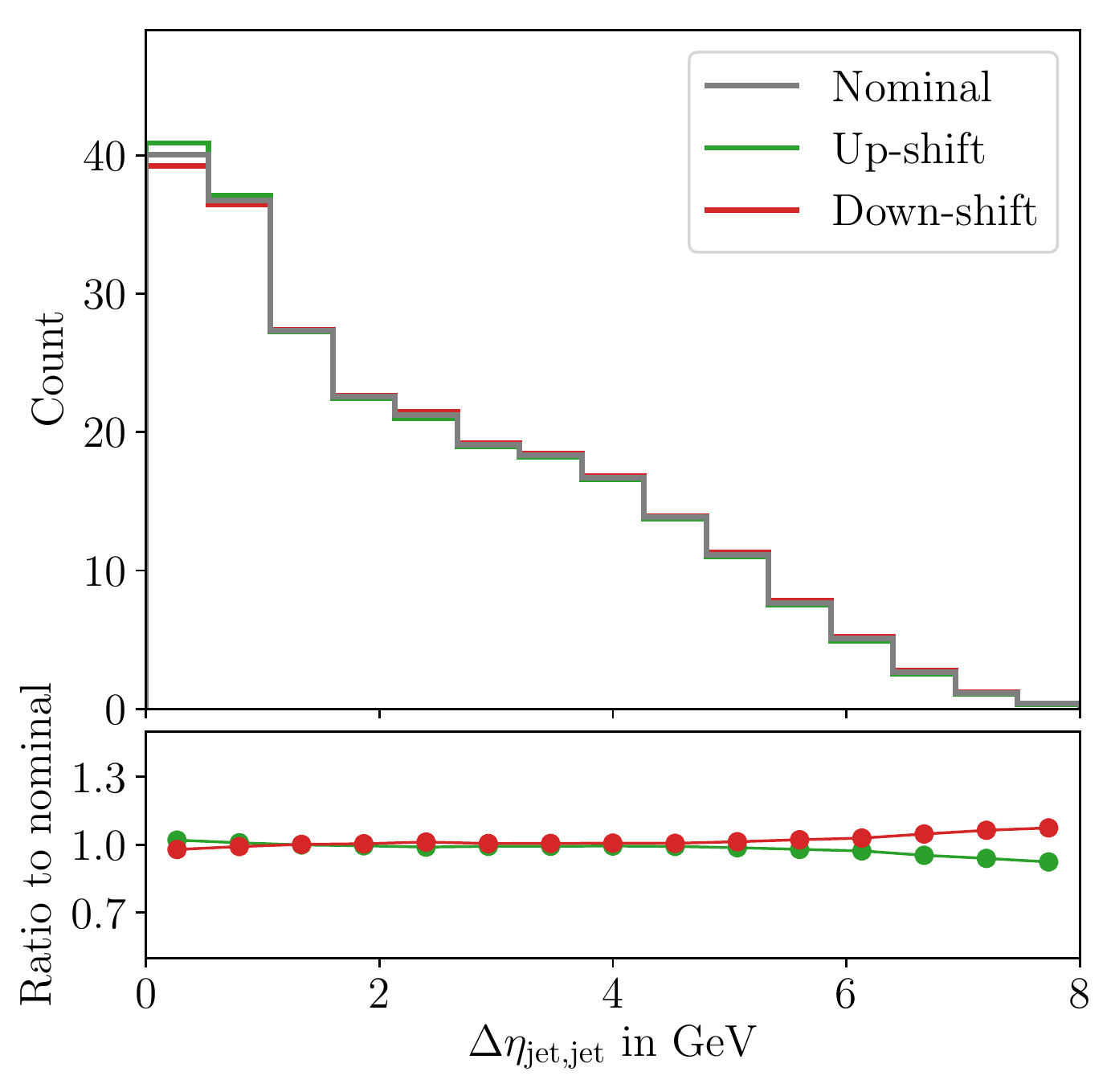}%
\includegraphics[width=0.5\linewidth]{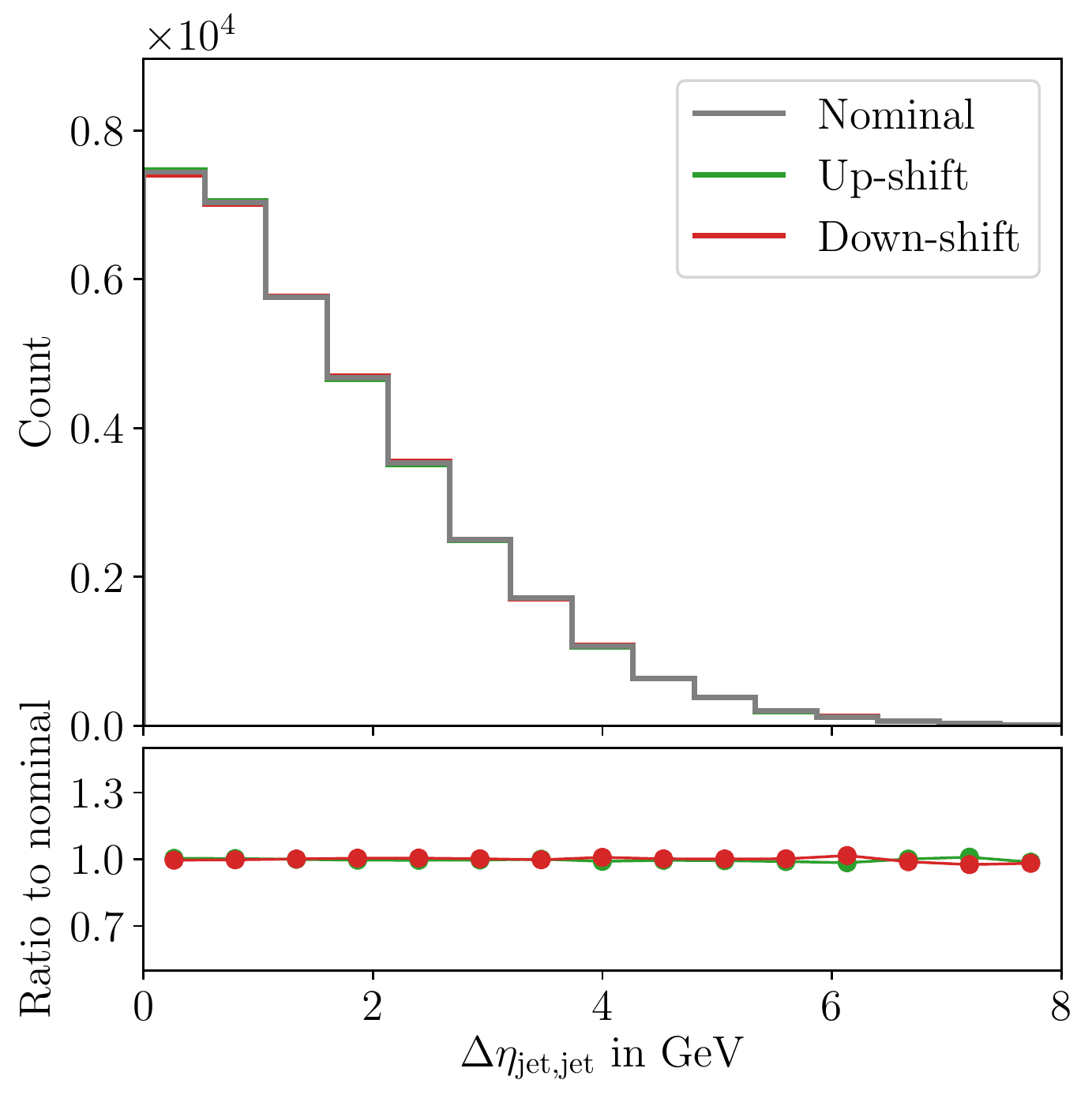}
\end{minipage}
\end{figure*}

\begin{figure*}[ph]
\centering
\includegraphics[width=0.34\linewidth]{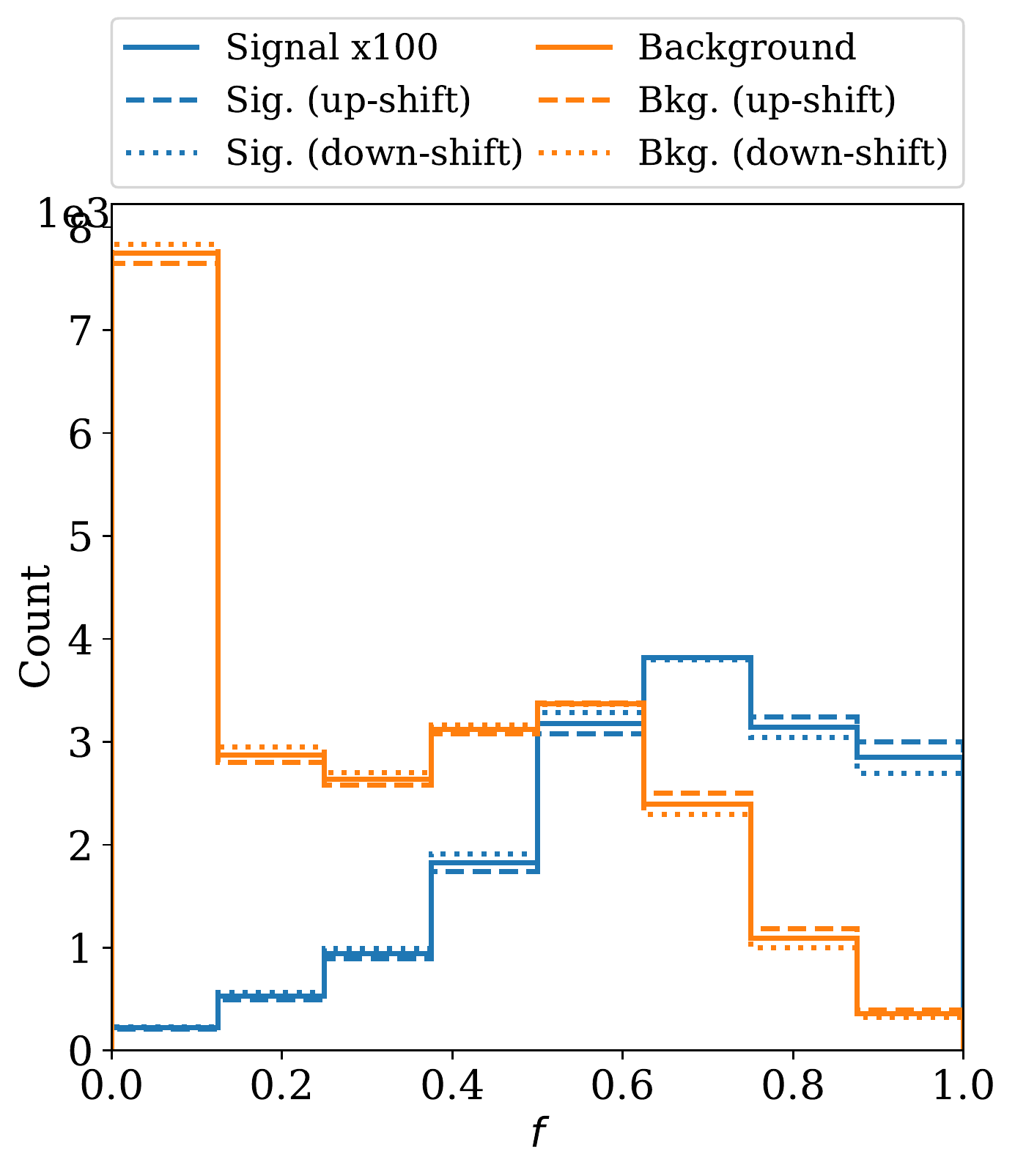}%
\includegraphics[width=0.34\linewidth]{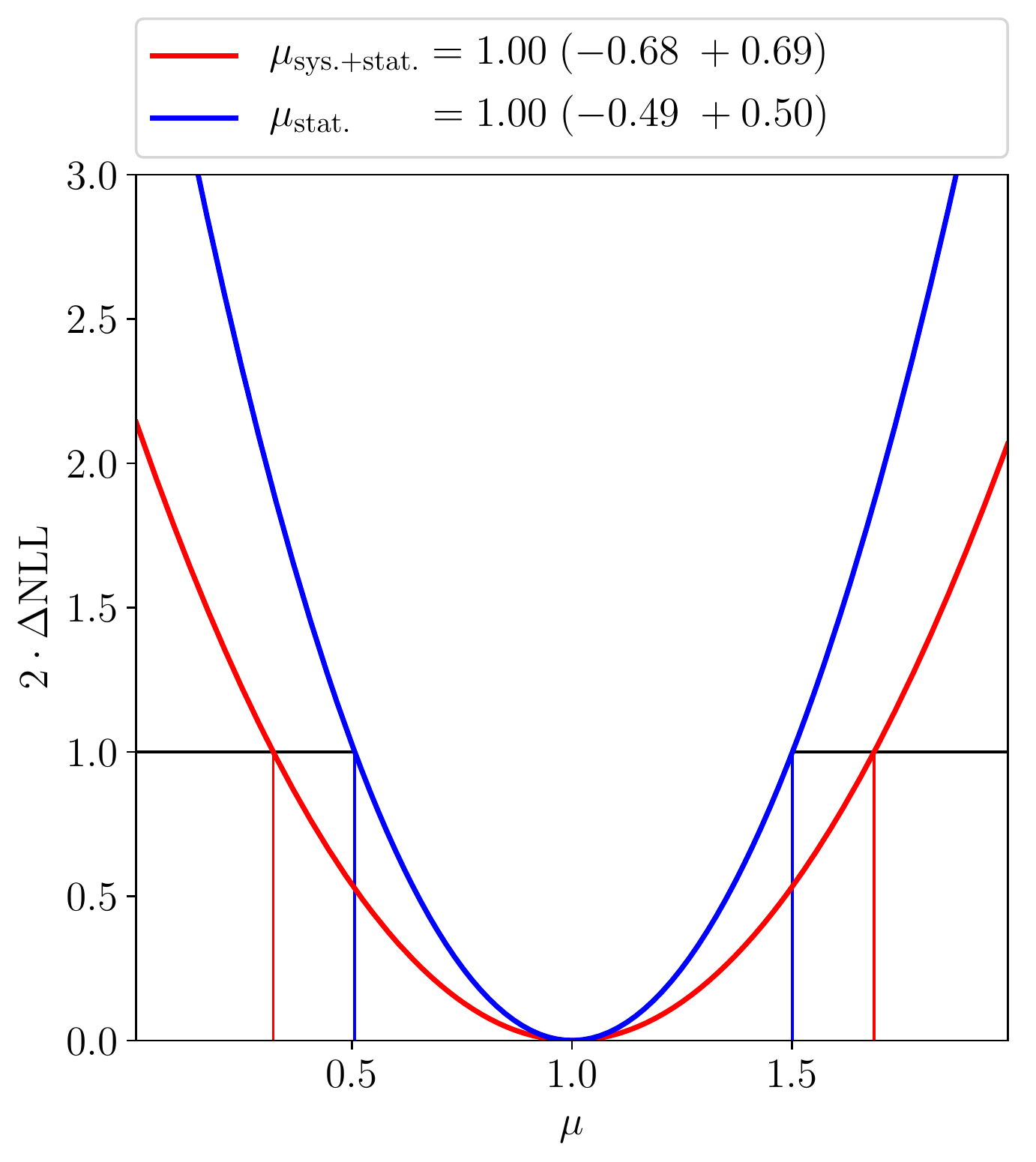}
\caption{Distribution of the \gls{NN} output for the more complex example of section~\ref{sec:Application_to_a_more_complex}, if the NN is trained on the classification of the two processes using the cross entropy loss. The likelihood profiles taking (red line) only the statistical uncertainty and (blue line) the statistical and systematic uncertainty into account for the final statistical inference of $\mu$ are shown on the right.}
\label{fig:higgs_ceonly}

\includegraphics[width=0.34\linewidth]{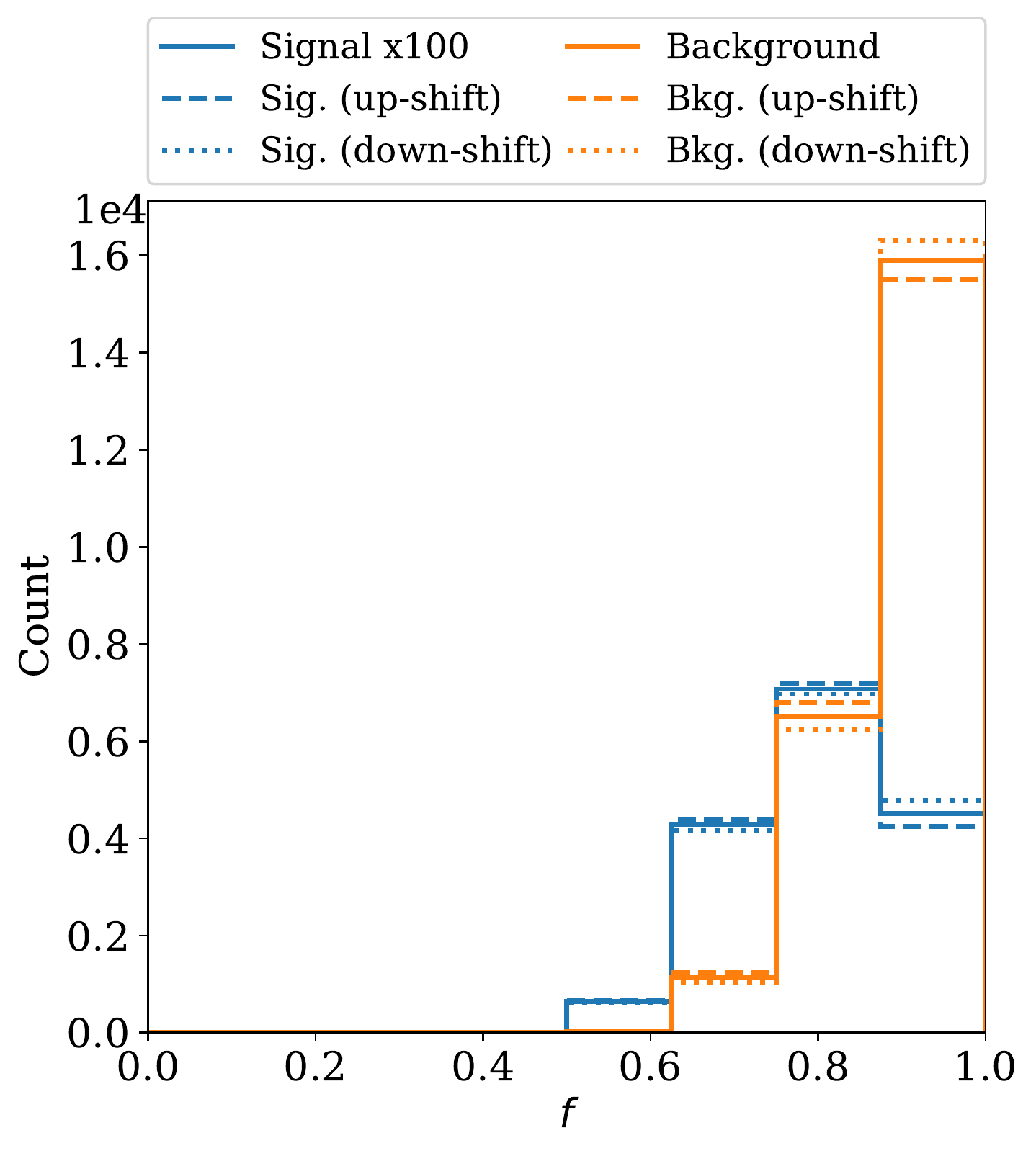}%
\includegraphics[width=0.34\linewidth]{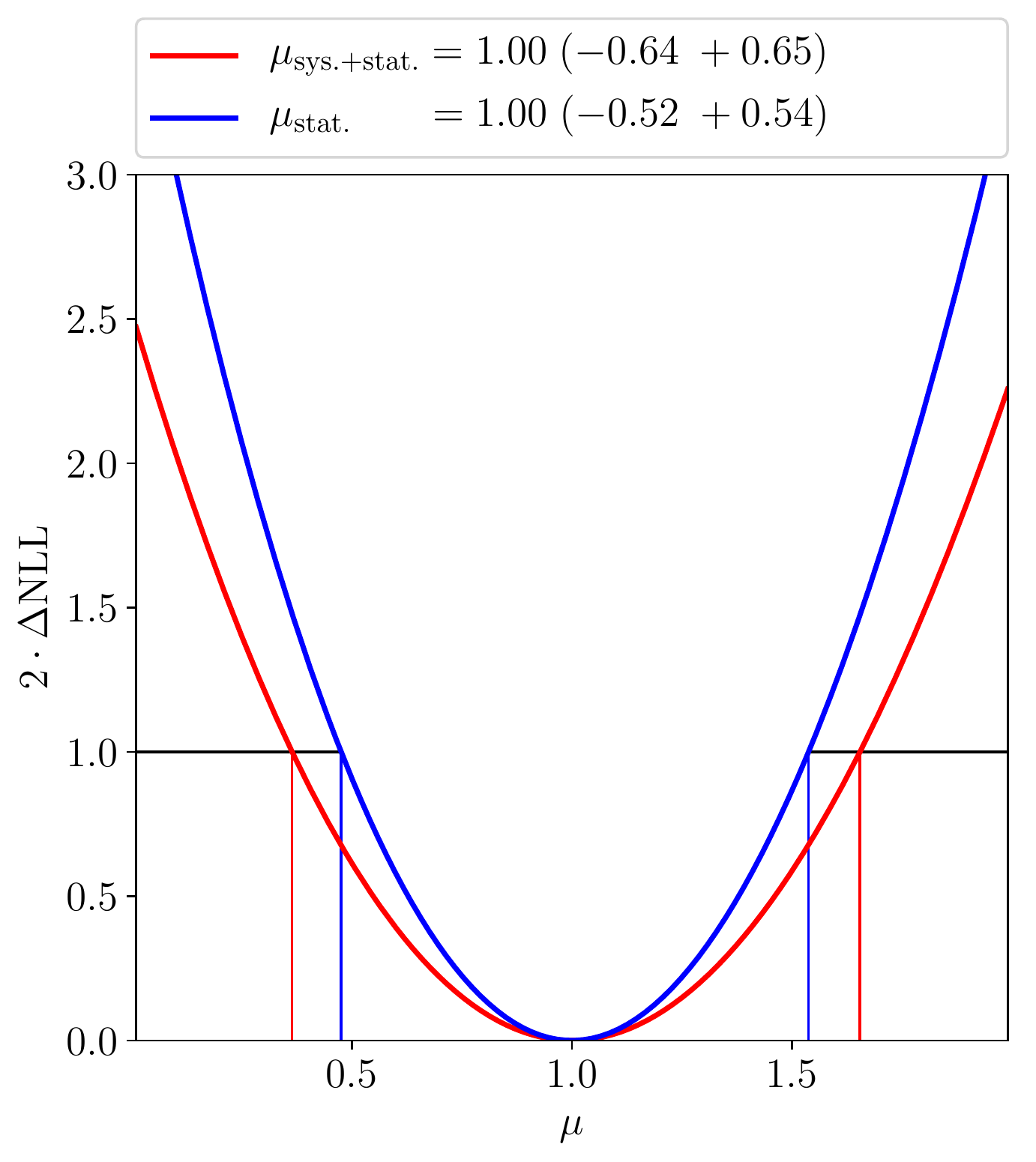}
\caption{Shown on the left is the distribution of the \gls{NN} output in the Higgs example for signal, background and the systematic variation if the NN is trained on the variance of the signal strength $V_{00}$ defined by the likelihood without the description of the systematic uncertainty. The likelihood profiles taking (red line) only the statistical uncertainty and (blue line) the statistical and systematic uncertainty into account for the final statistical inference of $\mu$ are shown on the right.}
\label{fig:higgs_statsonly}

\includegraphics[width=0.34\linewidth]{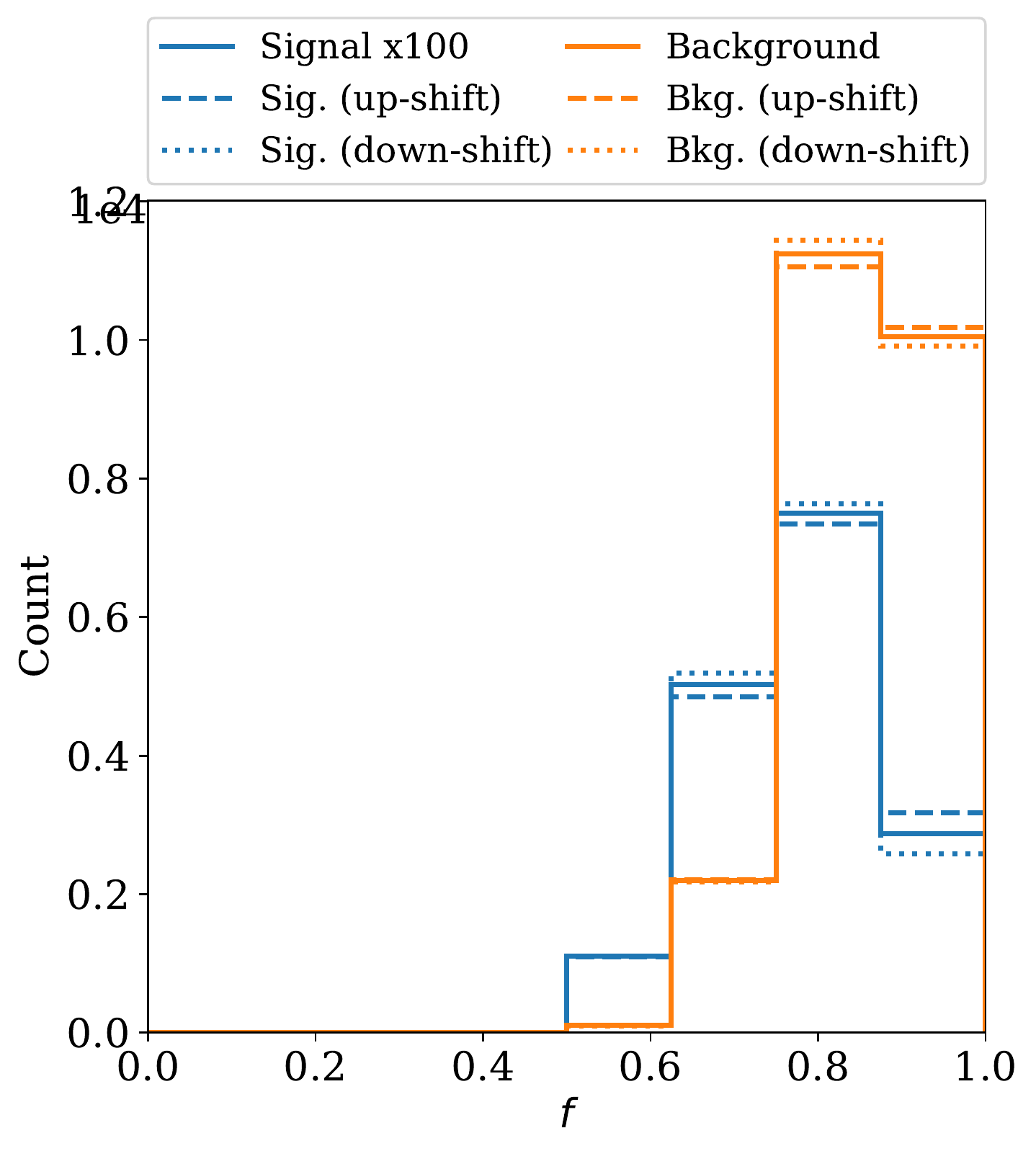}%
\includegraphics[width=0.34\linewidth]{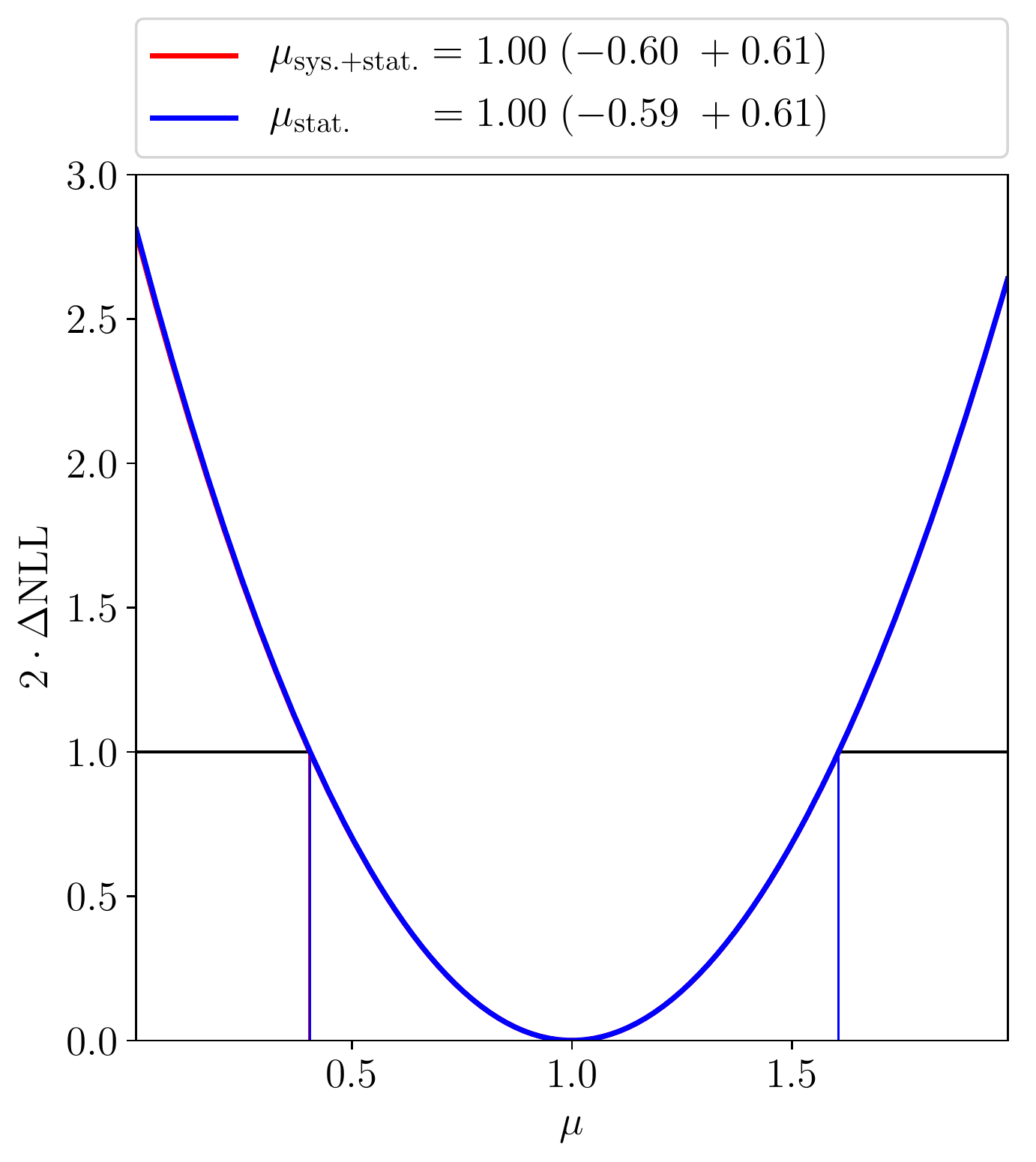}
\caption{Shown on the left is the distribution of the \gls{NN} output in the Higgs example for signal, background and the systematic variation if the NN is trained on the variance of the signal strength $V_{00}$ defined by the likelihood including the systematic uncertainty. The likelihood profiles taking (red line) only the statistical uncertainty and (blue line) the statistical and systematic uncertainty into account for the final statistical inference of $\mu$ are shown on the right.}
\label{fig:higgs_fullnll}
\end{figure*}

An (asymptotically) optimal result as derived for the previous example is not available since the likelihood in the input space is not known. Instead we use the training on the cross entropy loss as reference with $\mu=1.0^{+0.69}_{-0.68}$. Using $V_{00}$ as training objective, but without the implementation of the systematic variations of the input distributions in the loss function, the result for the signal strength $\mu=1.0^{+0.65}_{-0.64}$ shows a similar uncertainty compared to this reference. However, using the full likelihood from equation~\ref{eq:lh_full} as training objective, the signal strength is fitted with $\mu=1.0^{+0.61}_{-0.60}$. The inclusion of the systematic variations yields an improvement in terms of the uncertainty on $\mu$ of $\SI{12}{\percent}$ compared to the training on the cross entropy loss. The histograms and profiles of the likelihood used for extracting the results are shown in figures~\ref{fig:higgs_ceonly} to \ref{fig:higgs_fullnll}. For the assessment of the distributions of the \gls{NN} output, it should be noted that in contrast to the training based on the cross entropy loss, for the training based on $V_{00}$ no preference is given for signal (background) events to obtain values close to 1 (0). Similar to the result from the simple example in section~\ref{sec:Application_to_a_simple_example}, the profiles of the likelihood for all scenarios show that the training on $V_{00}$ removes the dependence on the systematic uncertainty yielding a smaller variance on $\mu$. On the other hand, the training on the cross entropy optimizes best the estimate of $\mu$ in the absence of systematic uncertainties, as expected from our previous discussion. With the proposed strategy, the \gls{NN} function learns to decorrelate against the systematic uncertainty, visible in the correlation of the signal strength $\mu$ to the parameter $\eta$ controlling the systematic variation, which drops from $\SI{69}{\percent}$ for the training on the cross entropy to $\SI{4}{\percent}$ for the training on the variance of the parameter of interest $V_{00}$, based on the full likelihood information as given in equation~\ref{eq:lh_full}.

\begin{figure*}[h]
\centering
\includegraphics[width=0.35\linewidth]{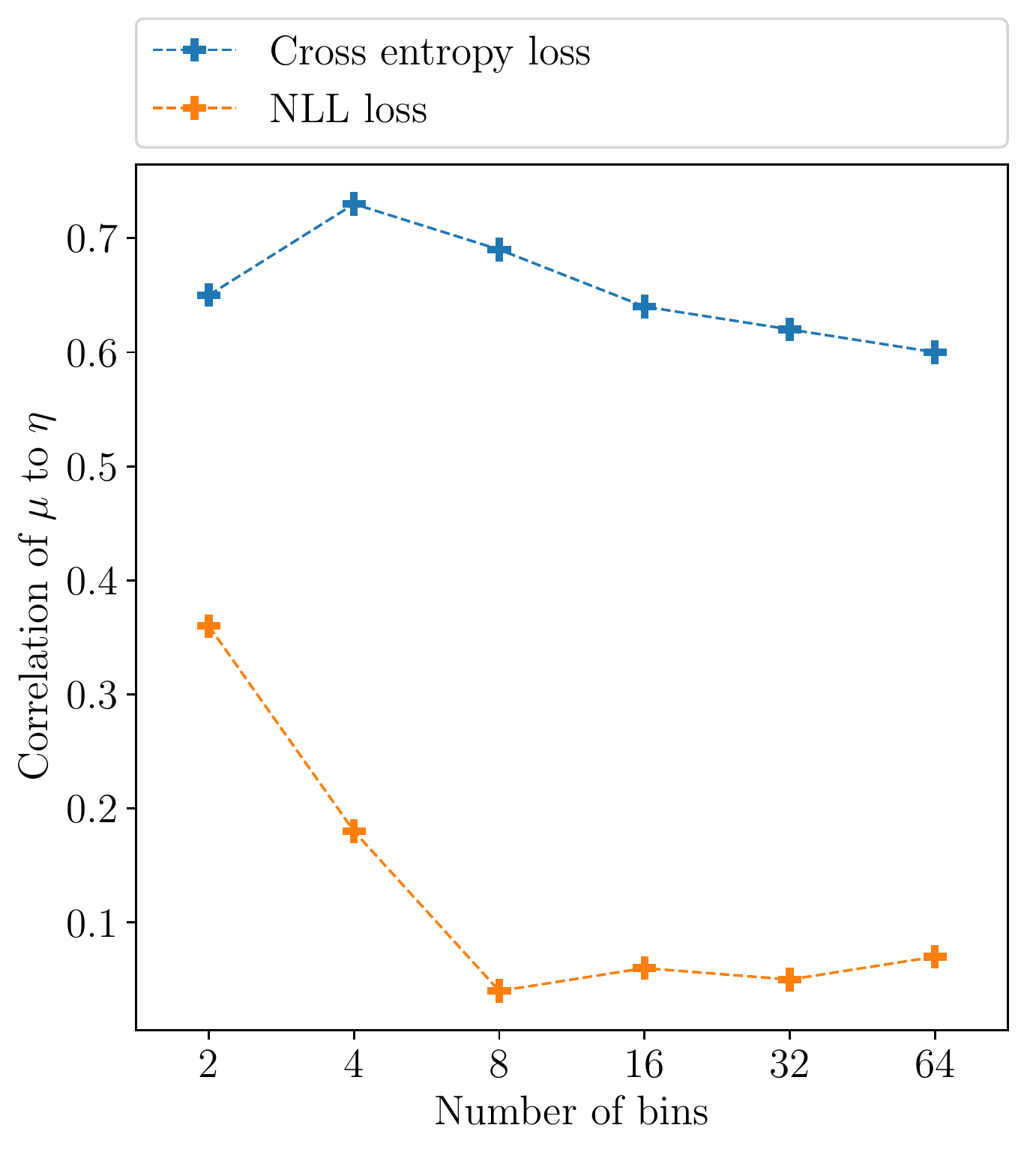}%
\includegraphics[width=0.36\linewidth]{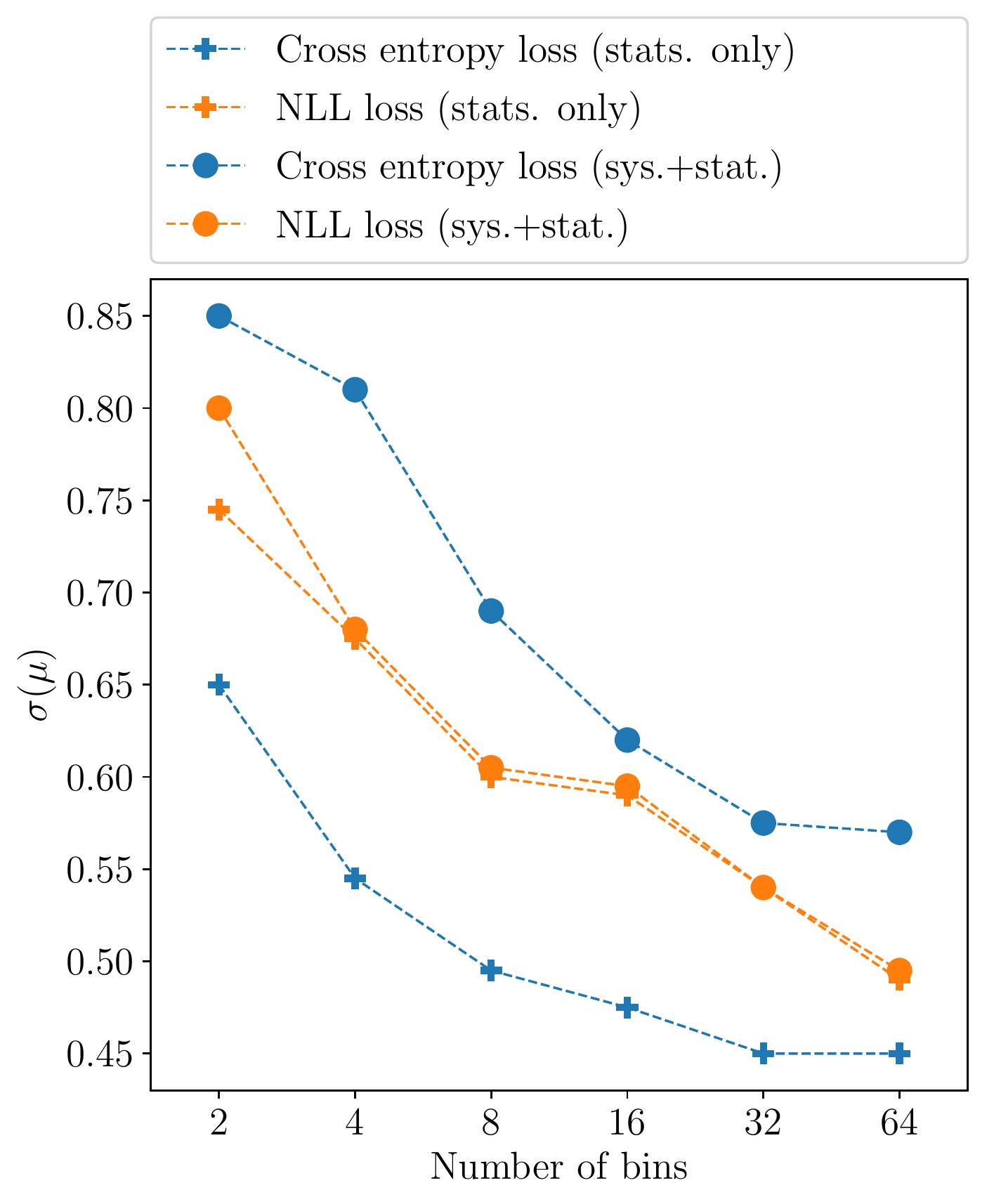}
\caption{Development of the (left) correlation between $\eta$ and $\mu$ and (right) the variance of $\mu$ ($\sigma(\mu)$) with the number of histogram bins for the training based on the cross entropy loss or $V_{00}$.}
\label{fig:higgs_nbins}
\end{figure*}

To improve the estimate of $\mu$ for the approach with the \gls{NN} trained on the cross entropy loss, a possible strategy could be to increase the number of histogram bins to exploit better the separation between the signal and background process. Figure~\ref{fig:higgs_nbins} shows the development of the performance with the number of bins for the training on the cross entropy loss and the training on the likelihood via $V_{00}$. The training on the cross entropy loss results in an estimate of $\mu$ with a mean correlation to the nuisance parameter $\eta$ of $\SI{66}{\percent}$ and a falling uncertainty in $\mu$ with an average distance of $0.18$ between the result for taking only the statistical uncertainties and statistical and systematic uncertainties into account for the statistical inference of $\mu$. In contrast, the strategy with the \gls{NN} trained on $V_{00}$ shows a reduction of the correlation between $\mu$ and $\eta$ of $0.35$ when moving from two to eight bins for the input histogram for the statistical inference. The estimate remains robust against the systematic variation for all tested configurations, yielding a smaller variance for the estimate of $\mu$ compared to the training on the cross entropy loss. The average distance between the inference using only the statistical part of the likelihood and the full statistical model is $0.01$. Including the systematic uncertainty in the inference, the comparison of the estimate of $\mu$ between the training based on $V_{00}$ and the training based on the cross entropy shows an improved variance of $\mu$ by 0.07 on average, yielding a stable average improvement of $\SI{10}{\percent}$.

It should be noted that in practice the granularity of the binning is limited by the statistics of data and the simulation. Limited statistical precision in the simulation is usually taken into account by introducing dedicated systematic uncertainties in the statistical model that typically degrade the performance of the analysis for a large number of bins.

\section{Summary}
\label{sec:Summary}

We have presented a novel approach to optimize statistical inference in the presence of systematic uncertainties, when using dimensionality reduction of the dataset and likelihoods based on Poisson statistics. Neural networks and in particular the differential approximation for the gradient of a histo\-gram enables us to optimize directly the variance of the estimate of the parameters of interest in consideration of the nuisance parameters representing the systematic uncertainties of the measurement. The proposed method yields an improved performance for data analysis influenced by systematic uncertainties in comparison to conventional strategies using classi\-fi\-ca\-tion-based objectives for the dimensionality reduction. The improvements are discussed using a simple example based on pseudo-ex\-pe\-ri\-ments with a known likelihood in the input space and we show that the technique is able to perform a statistical inference close to optimal by leveraging the given information about the systematic uncertainties. The applicability of the method for more complex analyses is demonstrated with an example typical for data analyses in high-energy particle physics. Future fields of studies are the application of the proposed method on analyses with many parameters in the statistical model and the evaluation of other possible differential approximations for the gradient of a histogram.

\section*{Acknowledgments}

We thank Lorenzo Moneta and Andrew Gilbert for helpful discussions and feedback, which greatly improved the manuscript.


\newpage
\bibliographystyle{splncs}
\bibliography{citations.bib}

\end{document}